\documentclass[useAMS,usenatbib]{mn2e}
\usepackage{deluxetable}
\usepackage[pdftex]{graphicx}
\usepackage[pdftex,usenames,dvipsnames]{xcolor}
\usepackage{amssymb}
\usepackage{natbib}
\usepackage{aas_macros}
\usepackage{url}
\usepackage[none]{hyphenat}
\usepackage{times}
\usepackage[linktocpage=true,backref=true,pagebackref=false,bookmarks=true,bookmarksnumbered=False,colorlinks=true,linkcolor=NavyBlue,citecolor=NavyBlue,urlcolor=NavyBlue]{hyperref}
\usepackage[all]{hypcap}
\pdfpageattr {/Group << /S /Transparency /I true /CS /DeviceRGB>>}  
\usepackage[backgroundcolor=yellow,linecolor=red,bordercolor=yellow,textsize=scriptsize]{todonotes}

\pdfoutput=1

\title[The central parsecs of M87]{The central parsecs of M87: jet emission and an elusive accretion
 disc\thanks{Based on VLT 4.B-0404(A), 76.B-0493(A), ALMA 2011.0.00754.5, NRAO AH 822, 862, 885, \textit{Chandra} 3088}}

\author[M.A. Prieto et al.]{M. A. Prieto$^{1,2}$\thanks{Email: aprieto@iac.es}, J.\,A. Fern\'andez-Ontiveros$^{3,4}$, S. Markoff$^5$, D. Espada$^{6,7}$, \newauthor O. Gonz\'alez-Mart\'in$^{8}$
\\
$^1$Instituto de Astrof\'isica de Canarias (IAC), E--38200 La Laguna, Tenerife, Spain.\\
$^2$Universidad de La Laguna, Dept. Astrof\'isica, E--38206 La Laguna, Tenerife, Spain\\
$^3$CEI Canarias -- Campus Atl\'antico Tricontinental (ULPGC--ULL), Universidad de La Laguna, Dept. Astrof\'isica, E--38206, Spain\\
$^4$Istituto di Astrofisica e Planetologia Spaziali (INAF--IAPS), Via Fosso del Cavaliere 100, I--00133 Roma, Italy\\
$^5$Anton Pannekoek Institute of Astronomy, University of Amsterdam, Science Park 904, 1098 XH Amsterdam, The Netherlands\\
$^6$Joint ALMA Observatory, Alonso de Cordova 3107, Vitacura, Santiago 763-0355, Chile \\
$^7$National Astronomical Observatory of Japan (NAOJ) and Department of Astronomical Science, The Graduate University for Advanced Studies \\ (SOKENDAI), 2-21-1 Osawa, Mitaka, Tokyo 181-8588, Japan \\
$^8$Instituto de Radioastronom\'ia y Astrof\'isica (UNAM), 3--72 (Xangari), 8701, Morelia, M\'exico}
\date{}

\pagerange{\pageref{firstpage}--\pageref{lastpage}} \pubyear{2016}

\def\LaTeX{L\kern-.36em\raise.3ex\hbox{a}\kern-.15em
T\kern-.1667em\lower.7ex\hbox{E}\kern-.125emX}

\begin{document}

\label{firstpage}
\maketitle

\begin{abstract}
We present the first simultaneous spectral energy distribution (SED) of M87 core at a scale of 0.4 arcsec ($\sim 32\, \rm{pc}$) across the electromagnetic spectrum. Two separate, quiescent, and active states are sampled that are characterized by a similar featureless SED of power-law form, and that are thus remarkably different from that of a canonical active galactic nuclei (AGN) or a radiatively inefficient accretion source. We show that the emission from a jet gives an excellent representation of the core of M87 core covering ten orders of magnitude in frequency for both the active and the quiescent phases. The inferred total jet power is, however, one to two orders of magnitude lower than the jet mechanical power reported in the literature. The maximum luminosity of a thin accretion disc allowed by the data yields an accretion rate of $< 6 \times 10^{-5}\, \rm{M_\odot \, yr^{-1}}$, assuming 10\% efficiency. This power suffices to explain M87 radiative luminosity at the jet-frame, it is however two to three order of magnitude below that required to account for the jet's kinetic power. The simplest explanation is variability, which requires the core power of M87 to have been two to three orders of magnitude higher in the last 200 yr. Alternatively, an extra source of power may derive from black hole spin. Based on the strict upper limit on the accretion rate, such spin power extraction requires an efficiency an order of magnitude higher than predicted from magnetohydrodynamic simulations, currently in the few hundred per cent range.
\end{abstract}


\section{Introduction}
Galaxies with core luminosities of about $10^{42}\, \rm{erg\, s^{-1}}$ fall at the borderline between nuclear starbursts and active galactic nuclei (AGN). When identified as AGN they often share a number of common features: extended radio lobes and often collimated jets, and an emission-line spectrum dominated by low-excitation lines often associated with an early-type galaxy. Usually, these low-luminosity AGN (LLAGN) are also referred to as LINERs \citep[low-ionization nuclear emission region][]{hec80}. The major challenge presented by these sources is that, despite their large supermassive black holes (SMBHs), the implied radiative efficiencies are orders of magnitude below the Eddington limit, assuming the standard $10\%$ mass-to-light conversion. M87 is a prime representative of this scenario \citep{ree82,fab95}. With a BH mass of $5.9 \times 10^{9}\, \rm{M_\odot}$ \citep[][re-scaled to $D = 16.4\, \rm{Mpc}$ adopted in this work]{geb09}, its core bolometric luminosity, $\sim 2.7 \times 10^{42}\, \rm{erg\,s^{-1}} $ (this work) implies an Eddington ratio of $ \sim 3.6 \times 10^{-6}\, L_{\rm edd}$, which is two orders of magnitude below that of a Seyfert nucleus (four below that of quasars with equivalent BH masses).

Several possible scenarios have been put forward to explain these low-luminosity nuclei: if the accretion rate is very low or the accretion mode inefficient, a decoupling between the temperature of the ions and the electrons may arise. This possibility led to the ion-supported tori models proposed by \citet{ree82}, to the advection-dominated accretion flow models \citep[ADAF;][]{nar94}, now generalized into several alternative radiatively inefficient accretion flows (RIAFs) models (e.g.\ ADIOS in \citealt{bla99}; CDAF in \citealt{qua00}), and to the jet-dominated models, in which the energy is predominantly released via relativistic outflows \citep[e.g.][]{fal04}. RIAFs became widely popular owing to their success in explaining the very low activity of Sgr~A$^*$ at the Galactic Centre and the low-hard state of BH X-ray binaries \citep[BHB; e.g.][]{nar95}. On the basis of Sgr~A$^*$ results, the RIAF solution has been extended to the general class of LLAGN, many of which harbour BH masses several orders of magnitude higher. To check on the validity of these or other types of accretion models requires well-sampled spectral energy distributions (SEDs) in both frequency and angular scales. Collecting this information is arduous for LLAGN because of their low luminosity and thus low contrast with respect to their host galaxy. Testing accretion physics in the context of low efficiency supermassive BHs, and in particular the suitability of RIAF models, is still reliant on poorly sampled SEDs in both frequency and spatial resolution even for the nearest and brightest LLAGN. Specifically, RIAF models predict a highly inverted spectrum in the radio band and multiple orders of inverse Compton scattering in the IR, optical, and X-ray ranges. The radio spectrum is typically well sampled, but the IR and optical bands are not, often relying on a few points, poor angular resolution, and upper limits. As a result, the core emission, buried in the host galaxy light, is unconstrained in the IR-to-optical range \citep[e.g.][]{rey96,mao08,era10,yu11}, an effect that translates into a degeneracy in the number of components and parameters of the models applied to these data.

Taking advantage of the high spatial resolutions accessible in the IR with large telescopes and adaptive optics, our group have been compiling the best possible sampled SEDs for a number of nearby LLAGN \citep{fer12}. Three key aspects are taken into account: the widest possible frequency coverage, sub-arcsec resolution across the entire spectral range, and data contemporaneity. Sub-arcsec resolutions --\,equivalent to physical scales between $10$ to $30\, \rm{pc}$ in our study sample\,-- minimizes the host galaxy contamination. Sampling the entire spectrum with the same aperture radius allows us a rigorous comparative analysis of the same physical region at all frequencies. Controlling variability --\,notorious in these sources\,-- allows us to isolate the source in a given energy state. The sample for which such an exercise is possible includes M87, NGC\,1052, NGC\,1097 and M104 (the Sombrero galaxy). The major outcome is that all the SEDs display a featureless spectrum, close to a power-law form, and are thus markedly different from those of Seyfert nuclei, obtained at comparable physical scales, and of quasars: they lack the blue bump feature due to emission from the accretion disc \citep[e.g.][]{ho97,mao07,era10} and the IR bump and the $1\, \rm{\micron}$ depletion, which are both fingerprints of the putative nuclear torus \citep{fer12}.

Specifically concerning the absence of the dusty torus, based on energetic arguments \citet{eli09} argue that the low luminosity of these sources makes them incapable of sustaining the vertical structure of a torus, the absence of strong nuclear winds would lead to a progressive flattening of any torus, in turn leaving a naked core. This progressive disappearance of the torus has indeed been observed with high angular resolution observations in H$_2$ molecular gas in our sample of nearby LLAGN \citep{mue13}. The torus also seems to be absent in FRI radio galaxies \citep{chi02} and in BL Lacertae sources \citep{plo12}, all these sources presenting instead commonalities with the LLAGN class. Still, not all LLAGN have a naked core, some present partial or total obscuration of their central source \citep{ho08,pri14}. Yet, as shown by the latter authors, this obscuration appears to be caused by large-scale (several $100\, \rm{pc}$) dust filaments crossing the nucleus, similar to the situation also seen in FRI radio galaxies \citep{chi02}.

This paper focuses on M87, its SED being the most illustrative case of a naked-core LLAGN. The large amount of data recorded for this object during the past twenty years allows us to handle its variability pattern across the electromagnetic spectrum in a robust manner. Two separate SEDs, each isolating a different activity state of M87, could be constructed. One SED represents the more regular semi-quiescent mode of M87; the second one represents an active --\,flare\,-- mode. For each of them, we assembled from the literature, archives, and our own observations, consistent SEDs in terms of angular scale (0.4 arcsec radius, projected $\sim 32\, \rm{pc}$) and frequency sampling (from the X-ray to the cm range, nine orders of magnitude in frequency). The use of high-angular resolution is specially critical in M87 because one of the brightest knots in its jet, HST-1, is $0\farcs85$ from the core. The nucleus of M87 is furthermore dust-free \citep[e.g.][]{pog02,mon14}, which allows us clear UV and X-ray detections. The spectra in these bands do not show the blue bump feature, nor is the Fe K$\alpha$ line seen in any of the longest exposure X-ray spectra taken of this source \citep[e.g.][]{mat03}, both being signatures of a standard thin accretion disc \citep{sha73}. M87 has been the subject of continuous testing of RIAF models with different flavours that also include the contribution of a truncated accretion disc and a jet \citep[e.g.][]{rey96,mat03,yu11,nem14,jon15}, although in all cases on the basis of a heterogeneous SED, using data with very different angular resolutions, limited wavelength sampling in particular in the IR and the millimetre regions, and no contemporaneity. The SEDs presented in this work are virtually free from these drawbacks (Section 2).

The optimal sampling of the compiled SEDs demonstrates the featureless nature of the M87 core spectrum, very different from the RIAF predictions, but closely resembling that produced by synchrotron jet emission (Section 3). Modelling of the SEDs on the basis of a jet plus a standard thin disc are presented in Section 4.

M87 shows a $20\, \rm{pc}$ nuclear rapidly rotating Keplerian disc in optical ionized gas almost normal to the jet direction \citep{har94,for94,mac97,wal13}. On the basis of the SEDs compiled in this work, no evidence for a counterpart of this ionized gas disc, or of the signature of a thin accretion disc or an advected geometrically thick flow are seen down to the finest scales marshalled in this work $12\, \rm{pc}$ from the core. On the basis of these findings, a strict upper limit to the accretion disc luminosity and mass accretion rate are provided and compared with literature estimates (Section 4.1).

A distance to M87 of $16.4\, \rm{Mpc}$ \citep[$1\arcsec = 80\, \rm{pc}$;][]{jor05} is assumed in this article.

\section{Extraction of the M87 core SED: angular resolution and variability}

M87 is one of the nearest massive elliptical galaxies. It has a $1.5\, \rm{kpc}$ jet seen from the radio to the X-ray region. In extracting the SED of the M87 core, we took into consideration a constant aperture radius across the entire spectrum in conjunction with data contemporaneity.

The jet of M87 spatially resolves into several knots. The most famous of these, HST-1, had a massive outburst in 2005, when the flux increased by factors up to 100 across the electromagnetic spectrum, surpassing that of the core by factors of up to three in the UV \citep{mad09} and 25 in the X-ray \citep{har09}. The proximity of HST-1 to the core, $0\farcs85$, and its brightness poses a problem with limited spatial resolution, in particular in the high-energy domain. The data used in this study at any spectral range and time period had had angular resolution of $0\farcs4$ \textsc{fwhm} or better, with the exception of the X-rays (see below). To make sure that the same physical scale is sampled at all frequencies, core fluxes in the two best-sampled SEDs compiled in this work are integrated in a fixed aperture radius of $r = 0\farcs4$ (projected $\sim 32\, \rm{pc}$). The compiled SEDs are listed in Tables 1 and 2, and flux extraction methods are described in Section 2.1. For comparison, an additional SED was extracted from the highest available resolution data. This SED combines VLBI radio, and {\textit{Hubble Space Telescope} (\textit{HST}) IR, optical and UV data. The integration aperture was fixed at a radius of $0\farcs15$ for the \textit{HST} data, to milli-arcsec for the VLBI data. The compiled SED is given in Table 4.

At high energies all the SEDs are restricted to \textit{Chandra} data only, and furthermore to measurements done when HST-1 was in a quiescent state. This restriction permits us a reasonable extraction of the core flux with minimum contamination from HST-1; at higher energies, the resolution is too poor, and further data beyond $10\, \rm{keV}$ are not used in the SEDs.

We note that SEDs based on datasets from aperture sizes $\gtrsim 1''$ would include a mixed contribution of different sources: the IR and optical ranges dominated by M87's bulge, and the radio to the millimetre by the jet \citep[see][]{fer12}. HST-1 would de facto be detectable at at all wavelengths from the radio to X-rays.

As regards variability, the core of M87 is variable on short and long timescales. At the longest timescales --\,months to years\,-- it shows a persistent variability with the higher amplitude changes, within a factor of 1.5--2 on scales of a few days at the highest frequencies --\,X-rays \citep{har09}. Towards the lower frequencies, the variability pattern shows a progressive decrease in amplitude with decreasing frequency: of at most $\sim 20\%$ in the UV, slightly lower in the optical, and undetected, within measurement errors in the IR and radio. This pattern is derived from a comparison of the following datasets. The \textit{HST}/UV light curve by \citet{mad09} shows flux changes of $\sim 15\%$ to $19\%$ with errors of $12\%$ during the quiescent period 1999--2004. The \textit{HST}/F606W optical measurements by \citet{per11}, in the quiescent period 2002--2003, show a maximum amplitude change of $19\%$ --\,reported errors of $1\%$\,-- whereas their \textit{HST}/UV measurements for the same period show flux amplitude changes dominated by the errors ($\sim 8\%$). The latter contrasts with Madrid's data in the UV, but since the period covered by Madrid is longer, seven years, the inferred $15$--$19\%$ variability is presumably more representative. In the IR, there are few available data: at $10\, \rm{\micron}$, the comparison of two measurements in the period 2000--2001 is consistent with no variability within the errors, $\lesssim 15\%$. In radio-cm, the comparison of the data collected in 2003 at $15$ and $8.4\, \rm{GHz}$ (this work) is consistent with no variability within the errors ($\sim 5\%$). These quiescent levels are consistent with those read from light curves in the 5--230 GHz range of \citet{had14} at any time outside the June 2012 M87 core outburst.

In addition, M87 further experiences occasional flares. To our knowledge, two of the best-monitored flares are the 1978 flare, detected at $2.3\, \rm{GHz}$ \citep{mor88}, and the 2005 flare monitored across the entire electromagnetic spectrum and coinciding with HST-1 outburst \citep{har09}. Further in time M87 outbursts, mainly triggered at gamma-ray energies and followed up in at least a lower energy range have subsequently been reported in the literature, but completion in energy range has never been so exhaustive as that of the 2005 outburst. For completeness, we quote some of the latest gamma-ray flares: the 2008 outburst, followed with VLBI and \textit{Chandra} \citep{acc09}, the April 2010 outburst followed with \textit{Chandra} and VLBA \citep{har11,abra12}, the March 2012 outburst followed with VLBI \citep{had14}.

Using as a reference the above flare periods, two SEDs of the core of M87 were extracted: a so-called quiescent state SED, which includes data acquired in periods outside the flare periods, and mostly from the year 2003, and an active state SED based on data exclusively collected during the 2005 flare. These SEDs are described in turn.

\begin{enumerate}
\item {\it The SED for the quiescent state} relies on data collected in 2003, the year with the largest set of contemporaneous data across the electromagnetic spectrum. This selection applies to the cm, optical, and UV range. Other ranges of the spectrum are sampled with data from years outside the above flare periods: the IR region with \textit{HST} data collected between 1998 and 2001, the millimeter region with ALMA data from 2012, the X-rays with the average of two \textit{Chandra} observations collected in 2000 and 2002. These two \textit{Chandra} observations are among the longest exposures; we used their average and an adopted error bar of 50\% to account for the persistent fast variability shown by M87's core in the X-ray region. The error is estimated as a variance of the average sustained flux over 9 years of monitoring, 2000 to 2008 \citep[fig.\,9 in][]{har09}. As regards the millimetre region, just at the time of the 2012 ALMA observations, M87 registered an outburst at $22$ and $43\, \rm{GHz}$ \citep{had14}. Examination of the light curves shown by these authors at $5$ and $230\, \rm{GHz}$ indicates no peak activity at these frequencies at that time. Moreover, we found that the trend defined by the ALMA data --\,which run over six independent frequencies\,-- joins the cm data smoothly from 2003 (also that from 2005, Fig.\,1). Thus, the ALMA data are include in this SED (also in the active SED, see below) but with an associated error bar of 40\%, which reflects the average flux increase seen at $22$ and $43\, \rm{GHz}$ in the 2012 outburst.

At higher energies the resolution is too poor; one could still expect the core to dominate at these energies. To verify this argument, the \textit{Suzaku}/XIS spectrum ($12$--$40\, \rm{keV}$) collected in Nov--Dec 2006 was extracted. It appears though that the spectrum is affected by a second HST-1 flare occurring at that time \citep{har09} and could not be used. The effect was confirmed by comparing the \textit{Suzaku} and \textit{Chandra} spectra: the spectral slopes in both datasets are similar, but \textit{Suzaku} detects an order of magnitude higher flux, presumably from HST-1.

\item {\it The SED for the active state} relies on radio-cm, optical and UV data collected in April--May 2005, and the near-IR data on January 2005. This SED is complemented with the 2012 ALMA millimetre data since the spectrum, as in the quiescent state, smoothly joins the radio-cm data of 2005. This effect is in line with the observed low variability in the cm range discussed above. In the X-ray region, the high luminosity of HST-1 during the 2005 flare hampers a reliable extraction of the core spectrum at that time. As the variability of the core keeps the same pattern and strength over the period 2000--2009 regardless of the flare \citep{har09}, we decided to use the same \textit{Chandra} average spectrum of the quiescent state and an error bar of 50\% also for the active phase.

The 2005 outburst was also observed in gamma-rays by H.E.S.S. \citep{aha06}. These authors report on a TeV short-timescale flaring at the time of HST-1 flare peak. Because of the fast variability, they concluded that the source of the TeV flare must be M87 core. We find that the time coincidence of the TeV flare with the maximum reached by HST-1 in any other range of the electromagnetic spectrum questions that identification, thus no further high energy data beyond the X-ray are considered in this SED.
\end{enumerate}

\subsection{Compiled fluxes in a nominal aperture radius of $0\farcs4$ ($\sim 32\, \rm{pc}$)}

Nuclear fluxes and errors for both active and quiescent states are listed in Tables 1 and 2 respectively, along with relevant information on the data sources. For both SEDs and all spectral ranges, fluxes are integrated in a fixed-aperture radius of $\sim 0\farcs4$.

Specific to the UV to IR spectral region where the underlying galaxy contribution is important, the stellar emission was removed by averaging the emission in circumnuclear rings taken a various distances from the core, and avoiding HST-1. The selection of the inner and outer ring radii was a compromise between being closer enough to the centre and avoiding the contribution of the PSF wings of the core and of HST-1 to these rings. The errors associated with the nuclear fluxes represent the uncertainty associated with the background estimate in different rings, and are in the 10--20\% range. Alternative methods to subtract the galaxy contribution by fitting, for example, a S\'ersic profile to the bulge were also examined. However, we find large residual errors up to 20--30\% in the fitting procedure in the central 0.2--0.3 arcsec radii due to imperfect matching of the PSF to the core in some \textit{HST} filters (PSFs were computed with the \textsc{TinyTim} software\footnote{\url{http://www.stsci.edu/hst/observatory/focus/TinyTim}}; \citealt{kri11}). This poor matching is caused by the persistent contribution of the inner jet knots, which are present at all wavelengths. It was thus decided to use integrated aperture photometry, which is the most conservative approach, the least prone to errors introduced by the modelling, and the easiest to apply in a systematic manner to all data used regardless of frequency. Aperture corrections were applied to the \textit{HST} data only. We follow prescriptions by \citet{sir05} for the ACS, \citet{hol95} for the WFPC2, and \citet{pro03} for STIS. Extinction correction due to our Galaxy was applied in both SEDs following the extinction curve ($R_V = 3.1$) and the extinction value ($A_V = 0.073$) given by \citet{sch98}. We find no evidence for additional dust extinction in M87 itself as inferred from the low $N_{\rm H}$ value derived from the \textit{Chandra} data analysis, the power-law form of the spectrum up to the UV, and the lack of evidence for dust in the centre of the galaxy as derived after the subtraction of the bulge light in different optical and IR images \citep{mon14}.

In the radio-cm, the reported fluxes are either average values of various measurements collected in the same year or just single measurements. The new data reported here were complemented with literature values if available at the suitable angular resolution and time period. Errors represent the average dispersion or a $5\%$ standard error, whichever is higher.
 
In the millimetre, the ALMA data collected in 2012 in bands 3, 6, 7, and 9 spanning the spectral range $93$--$630\, \rm{GHz}$, were used. In each band, four spectral regions were extracted from which images were constructed. M87 is unresolved in all bands, and thus fluxes and errors were measured on the central point-like source detected in all cases by integrating on the beam size associated with the angular resolution at each frequency. Beam sizes range from $0\farcs33 \times 0\farcs28$ at $635\, \rm{GHz}$ to $2\farcs7 \times 1\farcs5$ arcsec at $93\, \rm{GHz}$. These resolutions imply that HST-1's emission is de facto included at frequencies below $250\, \rm{GHz}$ but not at the higher ones. Yet, the trend shown by the ALMA dataset is very smooth (Fig.\,1), hinting that the contribution of HST-1 is unimportant.

In the X-ray region, \textit{Chandra} data were selected from observations taken with a minimum integration time of $0.4\, \rm{s}$ to avoid core saturation and pile-up effects. The datasets taken in July 2000 and July 2002 were selected because of their relatively high signal-to-noise. That of 2000 is published in \citet{mat03}, whose reported fluxes at $1$ and $10\, \rm{keV}$, corrected by the Galaxy H column density, are provided in Table 1. The 2002 dataset was processed by us; the fluxes at $1$ and $10\, \rm{keV}$ were derived after applying a power-law fit and corrected by the Galaxy H column density. The flux variation at $1$ and $10\, \rm{keV}$ between the two \textit{Chandra} observations is 20\%, whereas the nominal errors derived from the power-law plus H column density fit are $\sim 13\%$. Table 1 provides individual 1 and 10 keV fluxes for each epoch and their average. An error of 50\% is associated with these fluxes to account for the variability. These average values and errors are displayed in the SEDs in Fig.\,1.

As a reference, we further include in Table 1 for the quiescent SED the H.E.S.S. detection of June 2004, which is one of the lowest registered, as read from fig. 3 of \citet{aha06}. The authors report three further TeV detections around March 2004 and May 2006, and at the beginning of 2005. In 2005, the TeV flux was five times that of 2004, coinciding with HST-1 flare maximum, whereas those in 2004 and 2006 they are about a factor 2--3 lower than that in 2003. The 2005 detection should be mostly HST-1 but the other two are presumably M87's core: the amplitude change of factor two would be in line with that seen with \textit{Chandra} at softer energies.

\section{The shape of M87 core SED}

The SEDs for the quiescent and active states are shown in Fig.\,1 (blue and red, respectively). Besides the X-ray data, which are the same for both SEDs, the overall spectral shape is very similar in both states and can be formally described by a broken power-law: flux density, $S_{\nu} \propto \nu^{-\alpha}$. This prescription points to a compact jet as the origin of the core emission of M87. The main characteristics of the spectrum are \textit{i)} a flat radio spectrum; \textit{ii)} a turnover in the millimetre region, at about $150$--$200\, \rm{GHz}$, and a steep power-law slope ($\alpha \approx 1.1$) from the IR to the UV.

In amplitude, the active SED is shifted up in flux by about $1.5$--$2$ in the UV and optical, $\sim 1.3$--$1.1$ in the radio. The $\sim 1.5$--$2$ factor variability seen in X-rays --\,regardless of the epoch\,-- is captured in the UV/optical bands as the difference in flux amplitude between the quiescent and actives phases, but the time scales are very different: days in the X-ray vs two years at least in the UV and optical bands. In the IR range, the active SED is dimmer by a factor $1.4$ to $1.2$ than the quiescent SED; still, these factors are within the measurement errors. In the radio range, both SEDs show similar flux levels. This pattern of flux amplitude change with frequency, i.e. the largest at the highest frequencies, and no change at the lowest ones, is same as that seen during quiescent --\,i.e. outside flare periods (Section 2). Both SEDs may thus represent overlapping events in the routine life of M87, always marked by a persistent and relatively modest variability. The 2005 SED may represent the best isolated case of one of many, continuous activity peaks of the core. If the scale factor between both SEDs is interpreted as intermittent particle acceleration events at the jet-core, the observed pattern of decreasing flux amplitude with decreasing frequency should be expected; the higher energies will sample the fastest cooling particles and the most recent acceleration event, whereas the lower frequencies, where the cooling is longer, will sample the accumulated contribution of several acceleration events, all together leading to a flat radio spectrum with a mild flux variation.

\begin{figure*}[p] 
 \centering
 \includegraphics[width=0.8\textwidth]{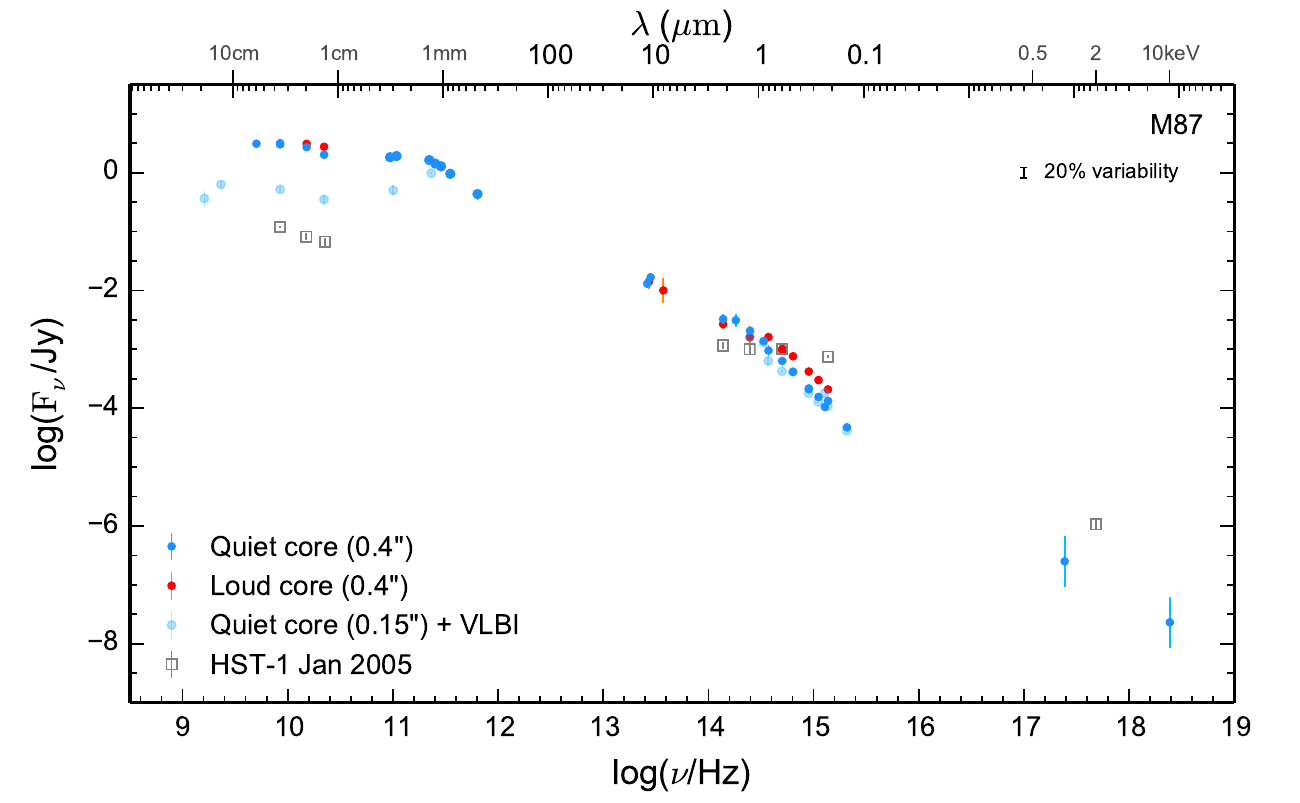}\\
 \includegraphics[width=0.8\textwidth]{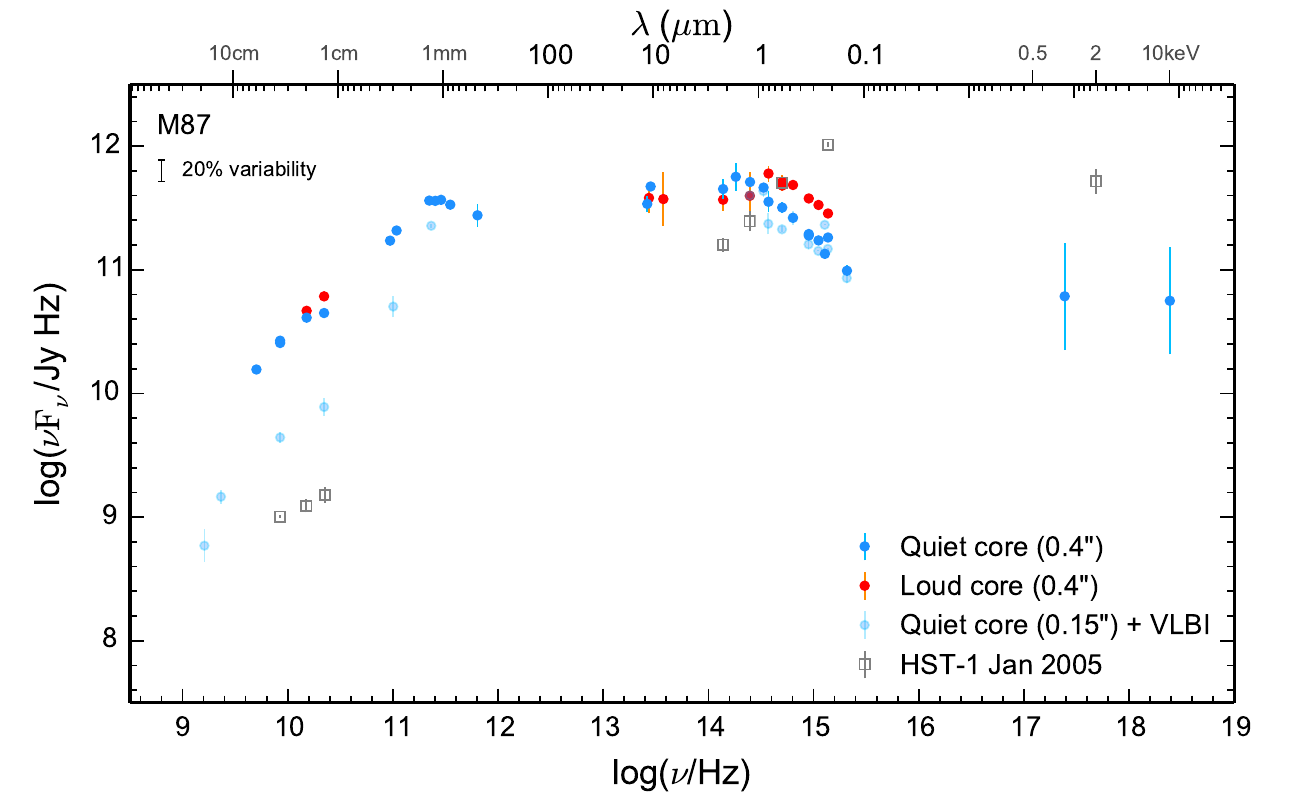}
\caption{SEDs of the central $0\farcs4$ ($32\, \rm{pc}$) radius of M87 (Table 1 and 2). Blue points represent the quiescent SED, red the active one, green that of the HST-1 jet knot. Light blue is the highest spatial resolution SED compiled in this work from a quiescent period: it combines VLBI data from milli-arcsec resolution and \textit{HST} data extracted from a $0\farcs15$ ($12\, \rm{pc}$) aperture radius. (a) SEDs shown in direct flux, (b) in power.} \label{fig1}
\end{figure*}

\subsection{Assessing the contribution of the jet knots in the $0\farcs4$ aperture radius}

M87 core fluxes within the $0\farcs4$ aperture radius include the contribution of the innermost jet-knots \citep{bir99,asa14,had14}. This section examines the relevance of this contribution in the core fluxes. First, we evaluate whether a re-acceleration event similar to that experienced by HST-1 may be dominating the emission in the active SED; second, we extracted the highest possible angular resolution SED of M87 core, within a radius of less than 0.15 arcsec ($\sim 12\, \rm{pc}$) and compared it with the nominal one of 0.4 arcsec radius.

For the first test, the SED of HST-1 during its flare peak in 2005 was extracted from a $0\farcs4$ aperture radius. The same datasets used to extract the active SED of the core were used. The procedure for extracting fluxes and errors are as described in Section 2, the SED is listed in Table 3. Specifically for the X-ray emission from HST-1, we used the $2\, \rm{keV}$ flux extracted by a special procedure in \citet{har06}. To cope with variability, this SED is strictly constrained to observations made in January--February 2005.

The SED is overplotted on those of the core in Fig.\,1, and no re-scaling is applied. Focusing on the direct-flux representation, it can be seen that the HST-1 SED departs markedly from those of the core. It shows a flatter spectrum in the optical--UV and is one order of magnitude brighter in the X-ray region. The spectrum is typical of those seen in hot spots in radio galaxies presenting an optical counterpart at the shortest wavelengths \citep[e.g.][]{mei97,pri97,mac09,ori12} whose emission is interpreted as very recent particle acceleration events caused by first order Fermi shock-acceleration. HST-1 in 2005 may just be experiencing an equivalent event. Conversely, M87 core SEDs present an inverted spectrum, i.e. a smooth flux decay with increasing frequency, which as a minimum tell us that none of the inner jet-knots are experiencing an outburst event equivalent to that of HST-1; hence, their contribution to our integrated core emission is not important.

For the second test, a third core SED using the highest angular resolution data available for M87 was constructed (Table 4, also shown in Fig.\,1). The SED combines VLBI at milli-arcsec resolutions or higher, and \textit{HST} data from the smallest feasible aperture size, i.e. commensurate with angular resolution and sampling of the \textit{HST} images. For that purpose, we used as a reference the \textit{HST}/UV image by \citet{bir99}, which shows the innermost resolved jet-knot, called \textit{L}, $0\farcs16$ from the centre. Accordingly, an aperture radius of $r \lesssim 0\farcs15$ ($\sim 12\, \rm{pc}$) was set for all the \textit{HST} images used for this SED. Aperture and extinction corrections were applied as in the case of the 0.4 arcsec SEDs. Both the VLBI and \textit{HST} data were selected from quiescent periods, and as contemporaneously as possible. The UV and optical data are all from the year 2003 and are the same as those used for the quiescent 0.4 arcsec SED. The VLBI and IR data span a large number of years, but, as discussed, the variability in these spectral ranges is found to be minor (Section 2).

As seen in Fig.\,1, the 0.4 arcsec and the new 0.15 arcsec SEDs are remarkably similar in shape and flux-levels from the UV to the IR despite a factor of seven difference in the aperture area. In the radio-cm, which VLBI samples at a few hundred Schwarzschild radii, the effect of the aperture size is, however, notorious. The VLBI-cm data, regardless of observation epoch, drops by a factor of six, yet the VLBI 1.3 mm has similar flux level to those sampled by the highest-frequency ALMA data of much inferior angular resolution. It can also be seen that the ALMA high frequencies do indeed mark a turnover in the 0.4 arcsec spectrum, becoming gradually flatter towards the lower frequencies (Fig.\,1). Using as further information the 5 and $1.6\, \rm{GHz}$ VLBI images by \citet{had14} and \citet{asa14}, the contribution of M87 jet into the 0.4 arcsec aperture SED can directly be assessed. At 5 GHz, the jet extends continuously out to a radius of 0.1 arcsec from the centre, at which point the emission suddenly falls off. At $1.6\, \rm{GHz}$, the jet extends further out to 0.4 arcsec and then falls off again. Putting all this information together, we can safely conclude that the dominant contribution in the 0.4 arcsec radius at the higher frequencies beyond 5 GHz is de facto the innermost 0.1 arcsec, $\sim 8\, \rm{pc}$, jet section seen at this frequency. If that is correct, the produced SEDs are delivering the most genuine representation of M87 core emission. 

We further interpret the spectrum turnover at about 1.3 mm as the transition frequency where the jet becomes progressively optically thick towards the lowest frequencies. On this basis, one may speculate that the UV to the mm band samples effectively the central few hundred Schwarzschild radii in both the 0.4 and 0.15 arcsec SEDs. Further interpretation of the VLBI-cm data, and of the highest angular resolution SED as a whole, in this scenario is complicated because of the dramatic VLBI drop in flux and the indication that this region is optically thick. All together this prevents a reliable association between the cm and the higher-frequency emission.Accordingly, the next sections focus on the modelling and analysis of the 0.4 arcsec aperture SEDs, which to our understanding provide the most coherent view of the central, presumably the inner $8\, \rm{pc}$, section of the jet-core across the entire electromagnetic spectrum. A comparative analysis with the highest angular resolution SED will also be discussed.

\subsection{Comparison with Seyfert and quasar SEDs at equivalent physical scales}

Fig.\,2 compares the M87 core SED (quiescent) with two different AGN templates: two Seyfert type 1 and type 2 templates derived from a sample of nearby Seyfert galaxies obtained at the same angular scale as that of M87 \citep{pri10} and a radio-loud quasar template from \citet{elv94}. Although the spatial resolution in the latter is in the arc-minute range in e.g.\ the IR or UV, we chose to represent the radio-loud class as it represents the most luminous quasars, and the core emission is expected to fully dominate their SED at any wavelength range.

\begin{figure*}[h] 
 \centering
 \includegraphics[width=0.8\textwidth]{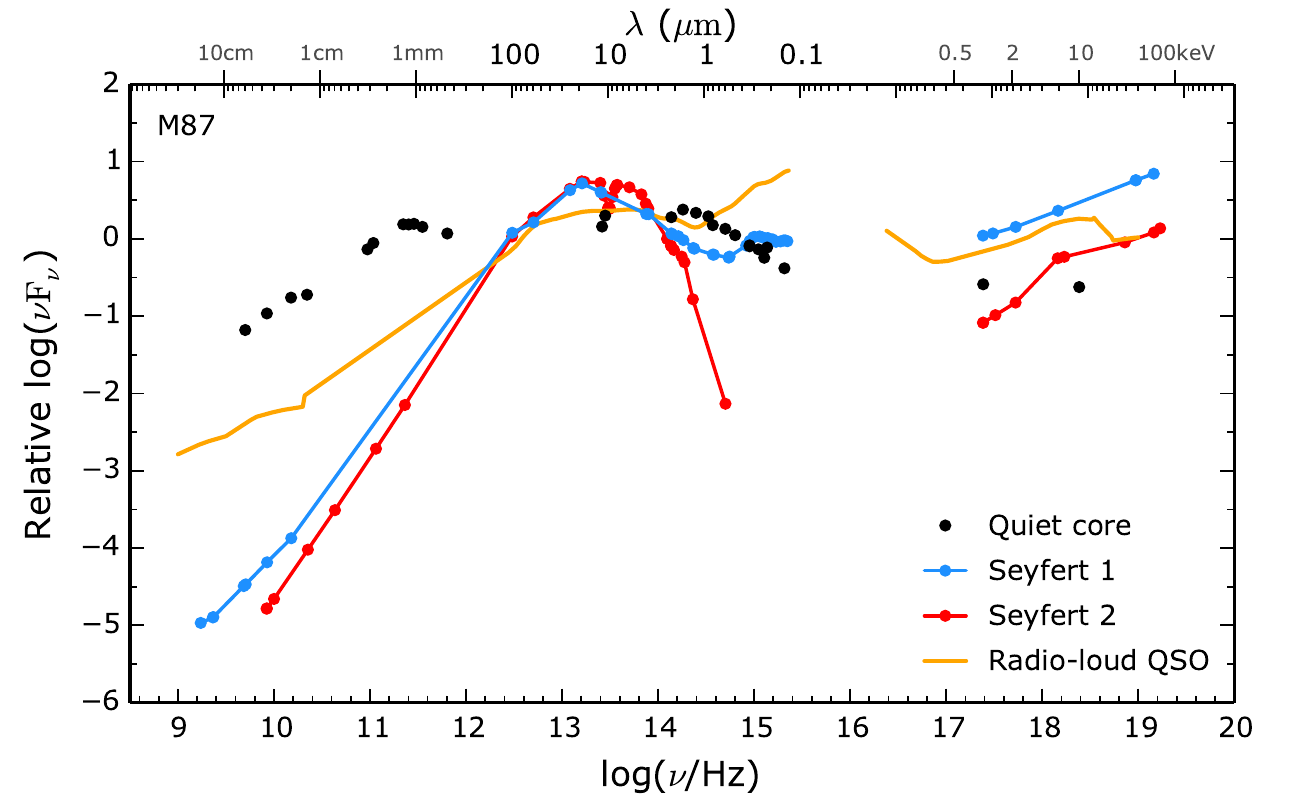}
 \caption{The M87 SED quiescent mode of the central $0\farcs4$ radius --\,black dots\,-- is compared with the Seyfert type 1 (blue) and type 2 (red) high spatial resolution SEDs from \citet{pri10}, which sample the central $\sim 10$--$15\, \rm{pc}$, and with the radio-loud quasar template (yellow) from \citet{elv94}. All the SEDs in the figure are normalized to the mean value of their $\nu F_{\nu}$ distribution (see text).}\label{fig2}
\end{figure*}

The sources in Fig.\,2 cover a wide range in luminosity, from $L_{\rm bol} \sim 10^{45-46}\, \rm{erg\, s^{-1}}$ in quasars, to $\sim 10^{43}\, \rm{erg\, s^{-1}}$ in the Seyfert class. M87 is at the lowest end with $L_{\rm bol} \sim 10^{42}\, \rm{erg\, s^{-1}}$. To facilitate the comparison, each SED is normalized to the mean value of the power integrated over the SED, i.e. the mean of the $\nu F_{\nu}$ distribution. As illustrated, Seyfert galaxies and quasars present characteristic common features in their SEDs whose relative strength marks the transition from completely obscured (Seyfert type 2), to fully unobscured (quasar) nuclei. Quasars and Seyfert type 1 show two prominent bumps in the optical/UV and the IR range associated with the accretion disc of the former and its reprocessed emission by dust the latter, yet Seyfert type 1 show still partial obscuration as their UV bump is relatively fainter and their near-IR emission below $2 \, \rm{\micron}$ --\,which traces the hottest dust\,-- is less prominent than in quasars. Both show the $1\, \rm{\micron}$ inflection point, that delimits the central dust sublimation radius. Conversely, Seyfert type 2 SEDs show the IR bump only, and a steep fall-off from the near-IR wavelengths onward, indicating that the hottest dust and the accretion disc are fully obscured in these sources \citep[see also][]{pri10}. In contrast, M87 SEDs lack all the above features, which suggests the absence of both a central obscuring dust structure and a standard thin accretion disc. Judging from the 100 ksec \textit{XMM} spectrum in \citet{gon09}, M87 also lacks clear evidence for a broad Fe K$\alpha$ line, a feature that is interpreted as the signature of cold material from a thin accretion disc. In the cm range, M87 is as loud as the average radio-loud quasar template. Thus, M87 SEDs may be the cleanest illustration of a jet core.

The featureless spectrum of M87 is also characteristic of other nearby LLAGN for which we have produced equivalent high angular resolution SEDs \citep[][and in preparation]{fer12}. The lack of the blue bump in LLAGN was already known in these sources \citep{ho99}; the additional absence of the IR bump and the $1\, \rm{\micron}$ inflection point (key components of the nuclear torus) is for the first time seen in our sample of LLAGN, of which M87 is the best illustration. The absence of all these features indicates that LLAGN are not scaled-down versions of the highly luminous ones. They may still represent a particular phase, and presumably the longest in time given the large fraction of LLAGN in local galaxies \citep{ho08}, in the AGN life cycle. It is interesting to note that FRI radio galaxies --\,M87 being one of those\,-- share most of the properties just described \citep{chi02}.

\section{Modelling M87 SED}

Because of the low luminosity of M87, its high BH mass, and in turn its low Eddington ratio, the source has received significant attention as a prototype of a radiatively inefficient source of the type seen in low BH mass X-ray binaries and Sgr~A$^*$ \citep{mat03,yua05,yu11}. It has also been explained as a scaled-down, otherwise normal AGN by \citet{mao07}.

As shown in previous sections, Fig.\,1 and 2, M87 SEDs are featureless. They depart from that of a standard AGN, they are also different from that inferred from RIAF models; specifically, they lack the multiple Comptoionization emission bumps characterizing these models (Section 1). On the contrary, M87 SED is reminiscent of a non-thermal spectrum. This section investigates whether a physical model based on a jet plus a thin disc can reproduce M87 SEDs. The disc is introduced on physical grounds: a disc is required as a reservoir of material to feed the hole and is an unambiguous component at high accretion rates. We introduce, however, a truncated disc to account for the absence of a blue bump.

The modelling follows the description by \citet{mar05,mar08} and \citet{mai09}. Briefly, the jet model consists of a superposition of multiple self-absorbed synchrotron components with a roughly conical distribution along its axis. This configuration results in a flat synchrotron spectrum at radio frequencies ($\alpha \gtrsim 0$, $S_\nu \propto \nu^{-\alpha}$), with a turnover frequency and an inverted spectrum at higher frequencies. The model used here assumes that the jet expands laterally with a constant sound speed, and the individual components in the relativistic plasma are weakly accelerated by the pressure gradient along the jet. A fraction of the initially thermal particles fed into the jets are accelerated into a power-law distribution and cool primarily via adiabatic expansion, while synchrotron and inverse Compton cooling determine the maximum energy of the particles.

\begin{figure*}[h] 
 \centering
  \includegraphics[width=0.67\textwidth]{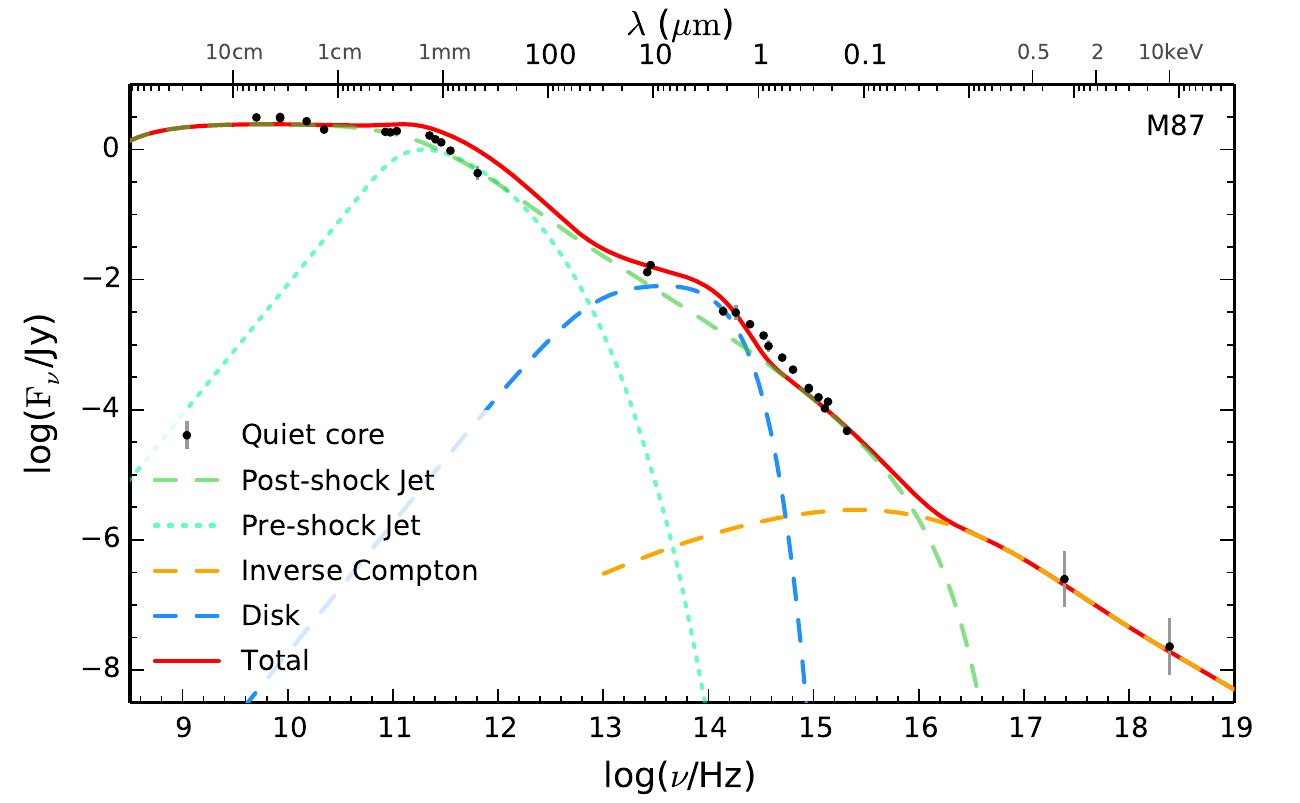}~
  \includegraphics[width=0.33\textwidth]{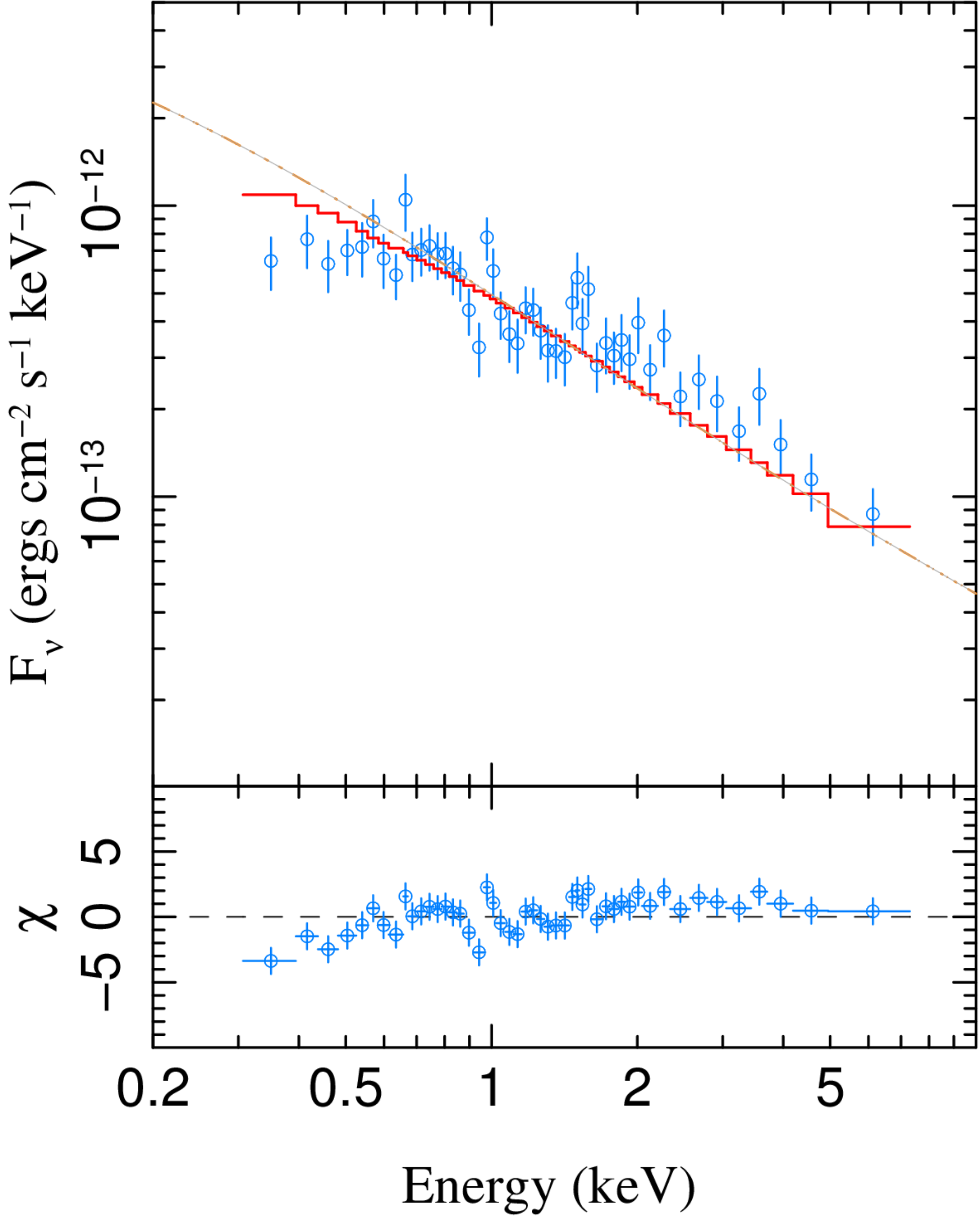}\\
  \includegraphics[width=0.67\textwidth]{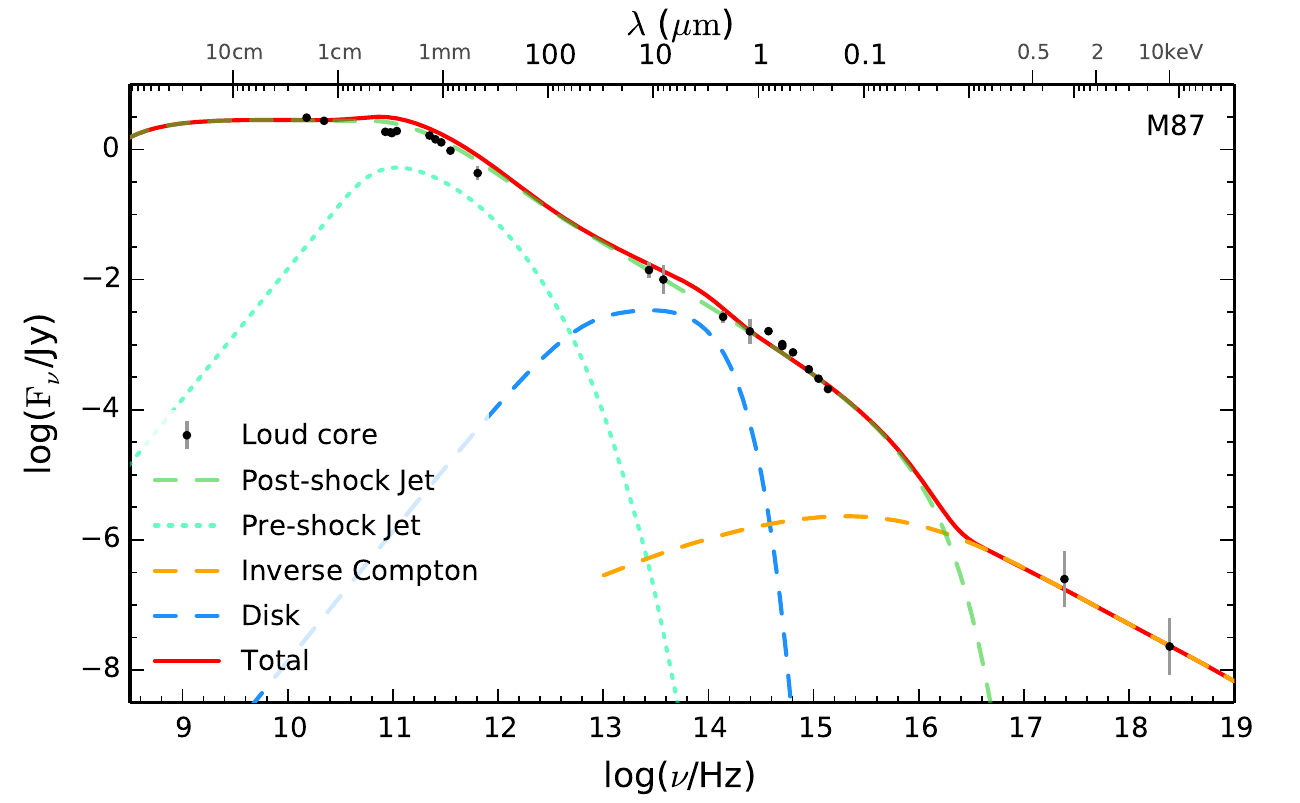}~
  \includegraphics[width=0.33\textwidth]{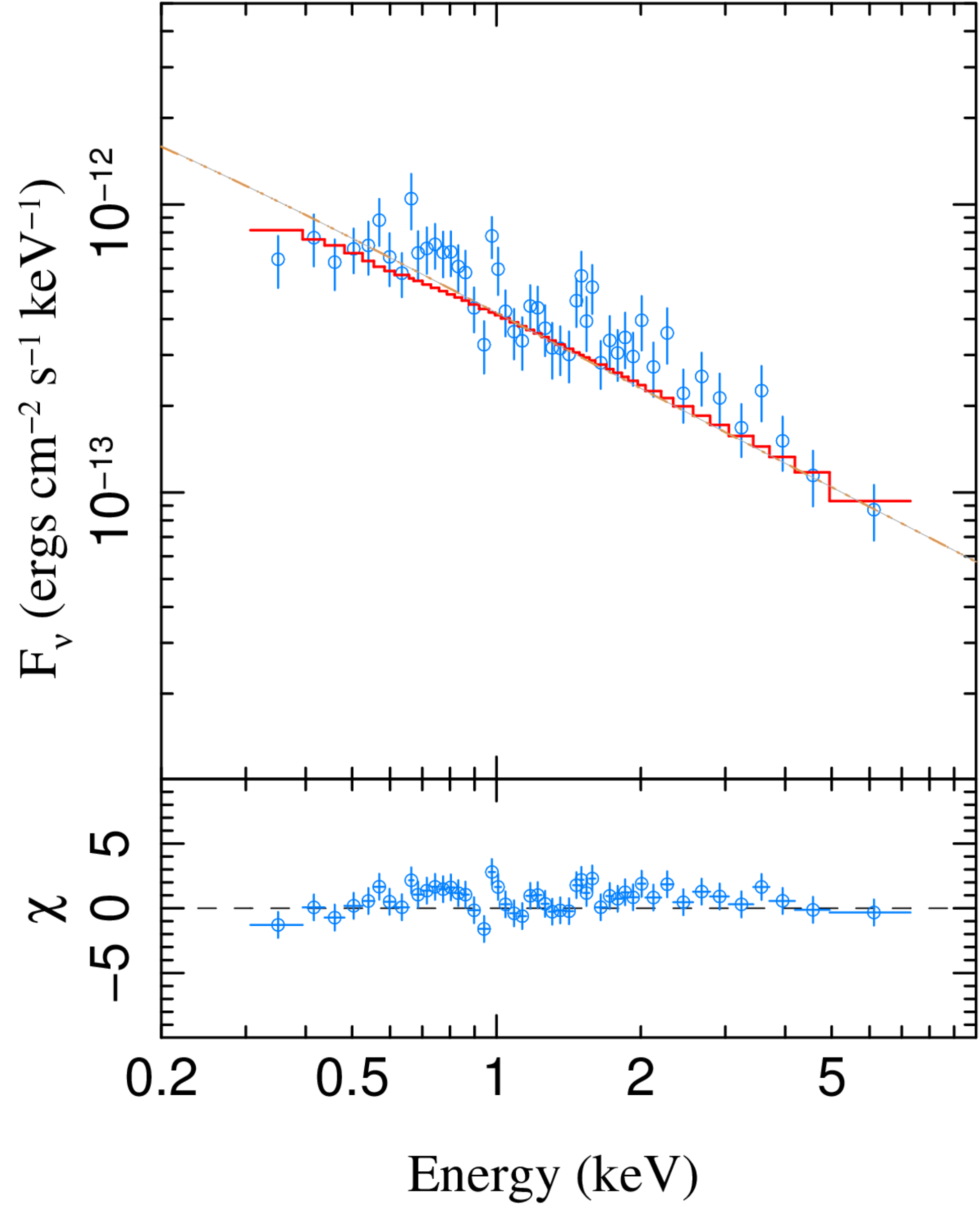}\\
  \caption{Jet--disc model result compared with M87 0.4 arcsec aperture radius ($32\, \rm{pc}$) SEDs. The top panel shows the result for the quiescent state; the bottom panel for the active one. The X-ray spectrum used in the model for both states is the 2002 \textit{Chandra} spectrum. The modelling of this spectrum is shown as a separate plot. Note, however, that the X-ray data plotted in the SEDs are an average of two \textit{Chandra} epochs, years 2000 and 2002, with error bars representing the average variability of M87 core in X-rays \citet{har09} --\,see text.} \label{fig3}
 \end{figure*}

\begin{figure*}[h]
  \centering
  \includegraphics[width=0.67\textwidth]{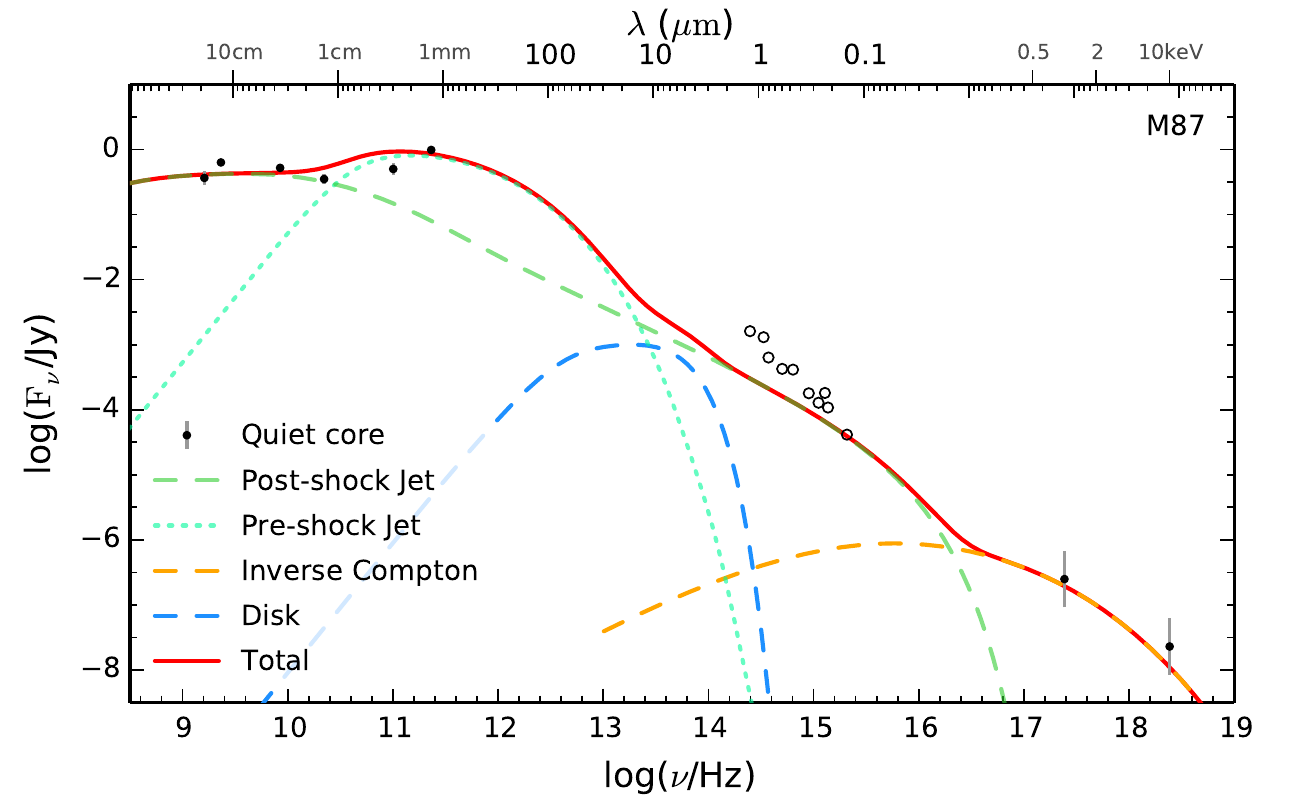}~
  \includegraphics[width=0.33\textwidth]{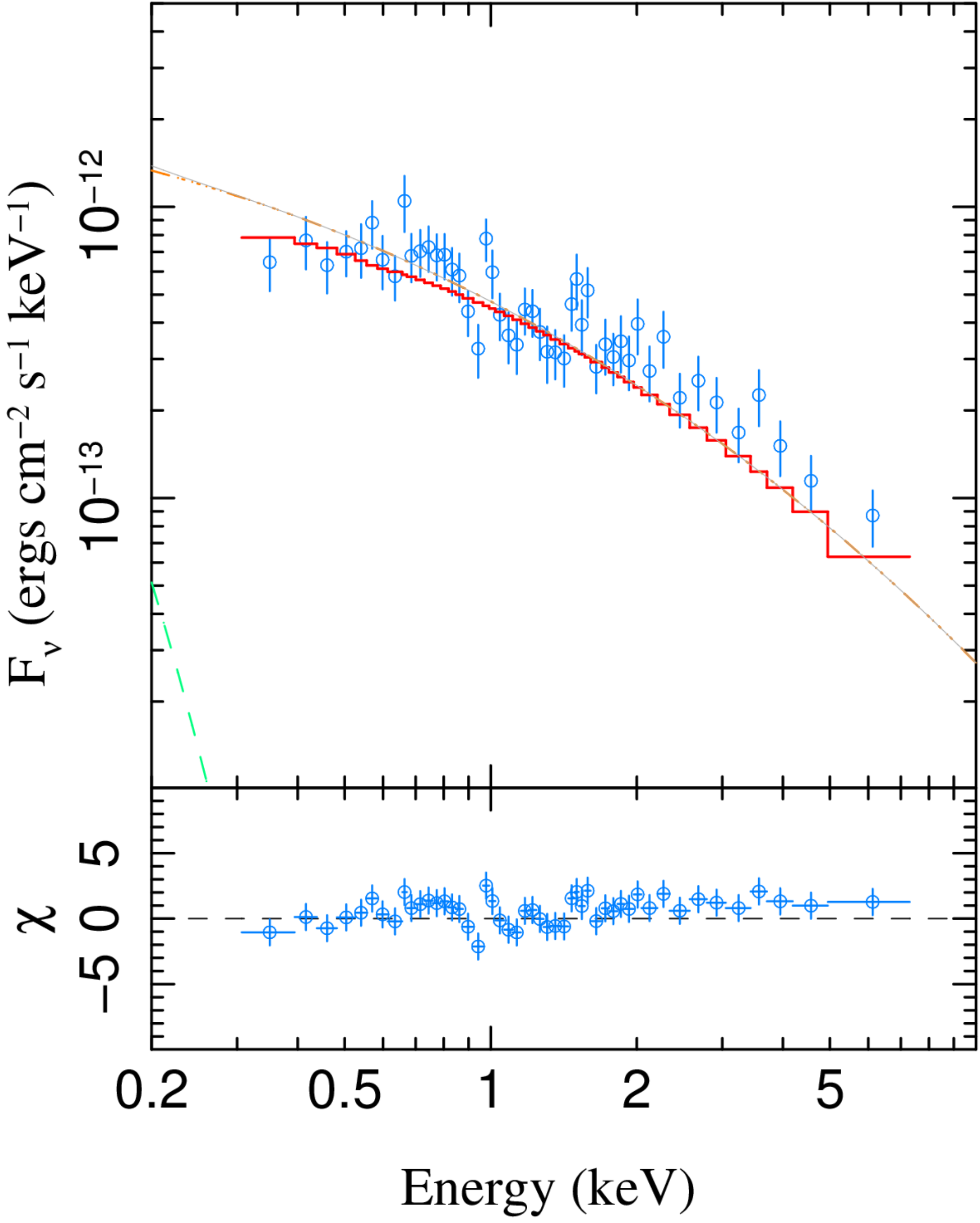}\\
  \caption{Jet--disc model for the highest angular resolution SED compiled in this work for a quiescent state. For the purposes of modelling, the optical-to-UV range, which is extracted from an aperture radius of 0.15 arcsec ($12\, \rm{pc}$), is taken as an upper limit to the core emission, and the power-law of the synchrotron spectrum is fixed at $p = 2.5$; the resulting model fit is based on the VLBI and X-ray data only, which are taken as representative of the core emission. The modelling of the X-ray spectrum --\,2002 \textit{Chandra} spectrum\,-- is shown in the small panel.}\label{fig4}
\end{figure*}

\begin{figure*}
  \centering
  \includegraphics[width=0.67\textwidth]{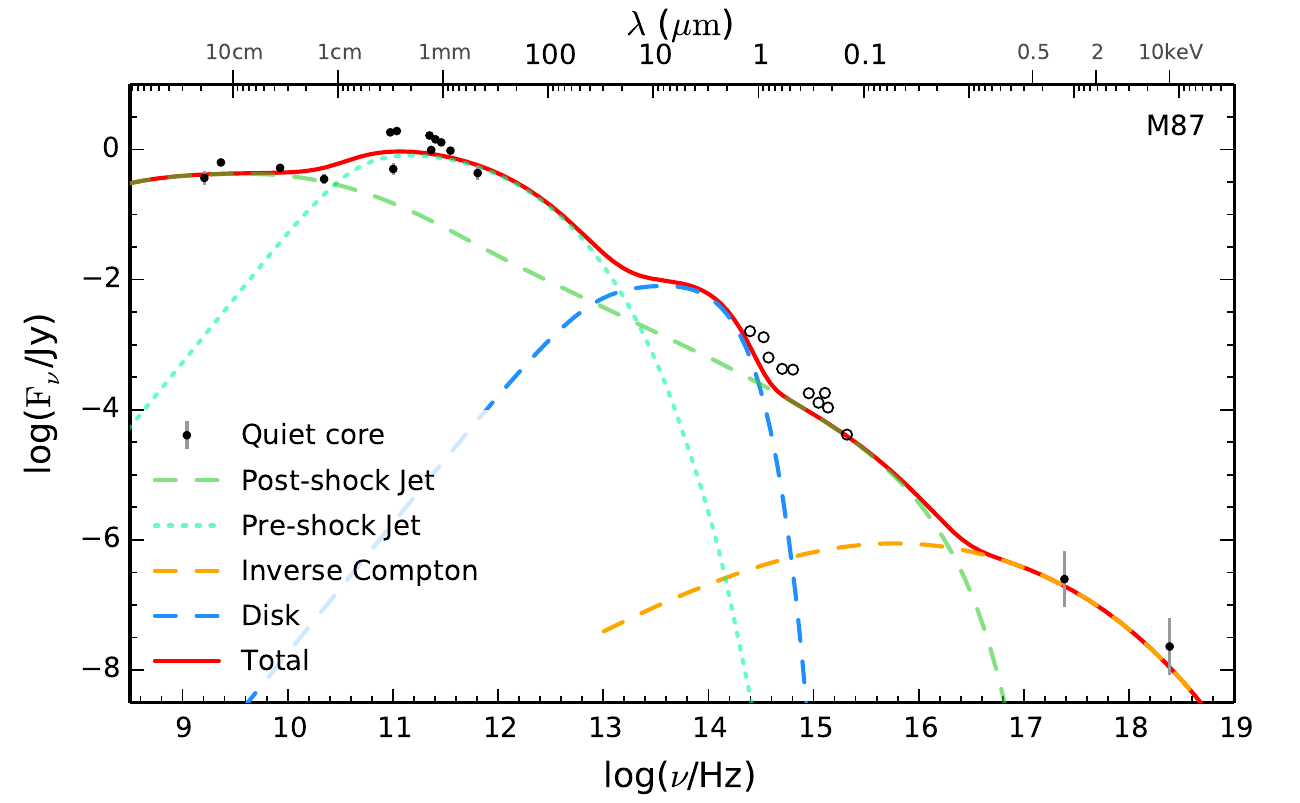}
  \caption{Jet--disc model for the highest angular resolution SED. The only difference with the model in Fig.\,4 is the contribution of the disc, which in this case is forced to be equal to that inferred from the 0.4 arcsec SED in quiescent mode. The fit of the \textit{Chandra} data is the same as that in Fig.\,4 and is not shown. For illustrative purposes, the ALMA data are added in this figure. They show that the emission from the highest ALMA frequencies reaches similar levels as the VLBI data at 1.2 mm, illustrating the transition from the optically thin to the optically thick region in the jet.}\label{fig5}
 \end{figure*}

In the model, the particles injected into the jet have their origin in a mildly relativistic and quasi-thermal plasma. This plasma is assumed to be advected into the jet nozzles from a radiatively inefficient accretion flow within the truncated disc. A multicolour blackbody accretion disc \citep{sha73} truncated at a certain radius is further included in the modelling. The SED is thus constructed out of direct jet synchrotron and inverse Compton emission components, together with the disc. The model is the same used to fit the hard state of BHB, Sgr~A$^*$ \citep[c.f.][]{fal00} and AGN like M81 or NGC\,4051 \citep{mar08,mai11}. The effects of relativistic beaming of the jets are taken into account on the basis of the inclination angle of M87 jet as determined from observations (see below). We note that an alternative model in which particles are first accelerated to a power-law distribution and then injected into the jet produces equivalent results to those derived from the assumption used here in which the acceleration occurs inside the jet after injection. M87's SEDs do not constrain any of these possibilities, yet, we found the latter more physically motivated.

The free parameters that enter into the model to characterize the jet are: the normalized input jet power ($N_{\rm j}$), the electron temperature of the relativistic thermal plasma entering at the jet base ($T_{\rm e}$), the energy partition factor $k$ expressing the ratio of magnetic to particle energy density, the physical dimensions at the jet base (radius $r_0$ and height $h_0$), the spectral index $p$ of the energy distribution for the accelerated electrons ($dN/dE \propto E^{-p}$), and the location at which particle acceleration starts ($z_{\rm acc}$). $N_{\rm j}$ corresponds to the total input power injected into particles and the magnetic field at the base of the jet. Since it scales with the total accretion power, it is expressed as a fraction of the Eddington luminosity. $r_{\rm in}$ and $T_{\rm in}$ are the radius and temperature at the inner edge of the disc, respectively. The model is most sensitive to the fitted parameter $N_{\rm j}$, which sets the power budget for the system.

A number of parameters derived from observations are fixed in the model. These are the equivalent H column density, $N_{\rm H} = 0.0194 \times 10^{22}\, \rm{cm^{-2}}$, as derived from a power-law fit to the July 2002 \textit{Chandra} spectrum. This value agrees with the Galactic value, in line with the results by \citet{mat03} after fitting the 2000 \textit{Chandra} spectrum; the distance to M87 \citep[$D = 16.4\, \rm{Mpc}$;][]{jor05}; the black-hole mass $M = 5.9 \times 10^{9}\, \rm{M_\odot}$ \citep[from][re-scaled to the above distance, hence $L_{\rm edd} = 7.4 \times 10^{47}\, \rm{erg\,s^{-1}}$]{geb09}; and the jet inclination of $15\, \rm{deg}$. The jet inclination angle is derived for the innermost jet section up to the HST-1 knot, and corresponds to the fastest proper motion measured in the M87 jet \citep{wan09}. Larger inclination angles are inferred from slower proper motions on jet knots further in distance. A likely range of values from 15 to 25 degrees is proposed in \citet{acc09}. To assess the impact on the model of a larger inclination angle, a model with a jet inclination of 25 degrees is also discussed below.

The X-ray data are fixed to the \textit{Chandra} observation collected in July 2002 in all the SEDs, regardless of angular resolution and M87 state (see Section 3.1 for reasoning). The model fits the \textit{Chandra} photon event spectrum after being folded through detector space using the program ISIS \citep{hou00}. The photon event spectrum was extracted within an aperture radius of $0\farcs5$ centred on the core of M87. Model fits to this \textit{Chandra} spectrum are shown in all cases in Fig.\,3. Moreover, to account for variability, the model result is compared with the average of the two \textit{Chandra} observations used in this work; namely, from the years 2000 and 2002. For that purpose, the average fluxes at $1$ and $10\, \rm{keV}$ from these two years are plotted in all the figures and compared with the model fit in Fig.\,3.

The parameters that are fixed in all the models are as follows: the fraction of accelerated electrons, set to 0.6, the outer radius of the accretion disc to $200\, R_{\rm g}$, gravitational radius, the maximum distance in the jet along which the electrons are accelerated, $z_{\rm max} = 10^{19}\, \rm{cm}$. We further assume that the energy density in protons and thermal pressure is the same. This assumption has a direct impact on the estimated $N_{\rm j}$; if assuming instead the number density in protons and electrons to be equal, $N_{\rm j}$ increases by an order of magnitude. We shall evaluate this assumption in the context of M87 results (Section 4.1).
 
All the models are set for a jet inclination of 15 degrees. Using instead an angle of 25 degrees yields an equivalent fit but one which, in the cm range, is underestimated. To compensate for the deficit, the total jet power, $N_{\rm j}$, has to be increased accordingly by 10--15\%. This change is a consequence of relativistic beaming of the jet, the larger the inclination angle, the more flux is beamed out of the line of sight, particularly at lower frequencies where the jet has the highest velocity at its base. As we are modelling the inner jet section, it was decided to use the angle of 15 degree in our analysis which appears to be the characteristic inner jet angle (see above).

Table 5 lists the resulting model parameters for both the quiescent and active SEDs. The values are similar for both SEDs. The model fit provides a reasonable account of both SEDs, and the relative contribution of the various components in the model are shown for each case in Fig.\,3. The jet component dominates the emission across the entire spectrum, the X-rays are the self-Comptonization of the jet synchrotron emission. The millimetre ALMA data is key in the modelling as it tightly constrains the location of the turnover frequency, which sets the transition from optically thin to optically thick emission (in the cm range). The disc contribution is minor. The quiescent SED shows a mild bump in the near IR at about $3\, \rm{\micron}$ which could be associated with a cold disc; the active SED shows no sign of such a bump, which is somehow unexpected if the active mode were interpreted as a re-activation of the accretion disc, as seen, for example, in the spectral changes shown in BHB. The mild bump could also be interpreted as a signature of an incipient acceleration event at one of the jet knots.

For comparison, the modelling of the highest angular resolution (0.15 arcsec radius) SED is shown in Fig.\,4. In this case, the model fit relies on the VLBI and X-ray data alone, on the premises that both spectral ranges sample the innermost core-jet region, whereas the optical and UV data are taken as upper limits to jet-core emission. To constrain the model, the spectral index $p$ was fixed to $p = 2.5$, consistent with a weakly efficient shock acceleration. The model parameters are listed in Table 6. The resulting model is less robust than in previous cases owing to the effectively poor sampling of the SED. The model still fits correctly the cm to mm and X-ray data with the jet component alone, the disc contribution being minor and in any case unconstrained because of the assumed upper limits in the optical--UV range. This spectral region therefore appears as an emission excess over the jet component.

Because of the similar flux level and shape of the 0.4 and the 0.15 arcsec SEDs in the optical--UV range (Fig.\,1), a second model in which the disc contribution is fixed to that inferred for the 0.4 arcsec quiescent SED was attempted. In this model, the location of the disc peak is left free, and the optical--UV data are still kept as upper limits. The result, in Fig.\,5, provides a much better fit of the entire SED, although the spectral shape in the optical--UV, closer to a power-law form, is not well reproduced. If the optical--UV data are taken as input to the model, i.e.\ not as upper limits, the resulting fit is similar (not shown).

Fig.\,6 shows the VLT-NACO $2\, \rm{\micron}$ image of the central kpc of M87. The angular resolution of the NACO data used in this work, $\sim 0.15$ arcsec \textsc{fwhm}, allows the separation of HST-1 emission from that of the core in both the quiescent and active phases of M87. At the time of this observation, M87 was at the peak of its 2005 outburst, and the HST-1 flux at $2\, \rm{\micron}$ was half the flux of the core.

\begin{figure*}
  \centering
  \includegraphics[width=0.67\textwidth]{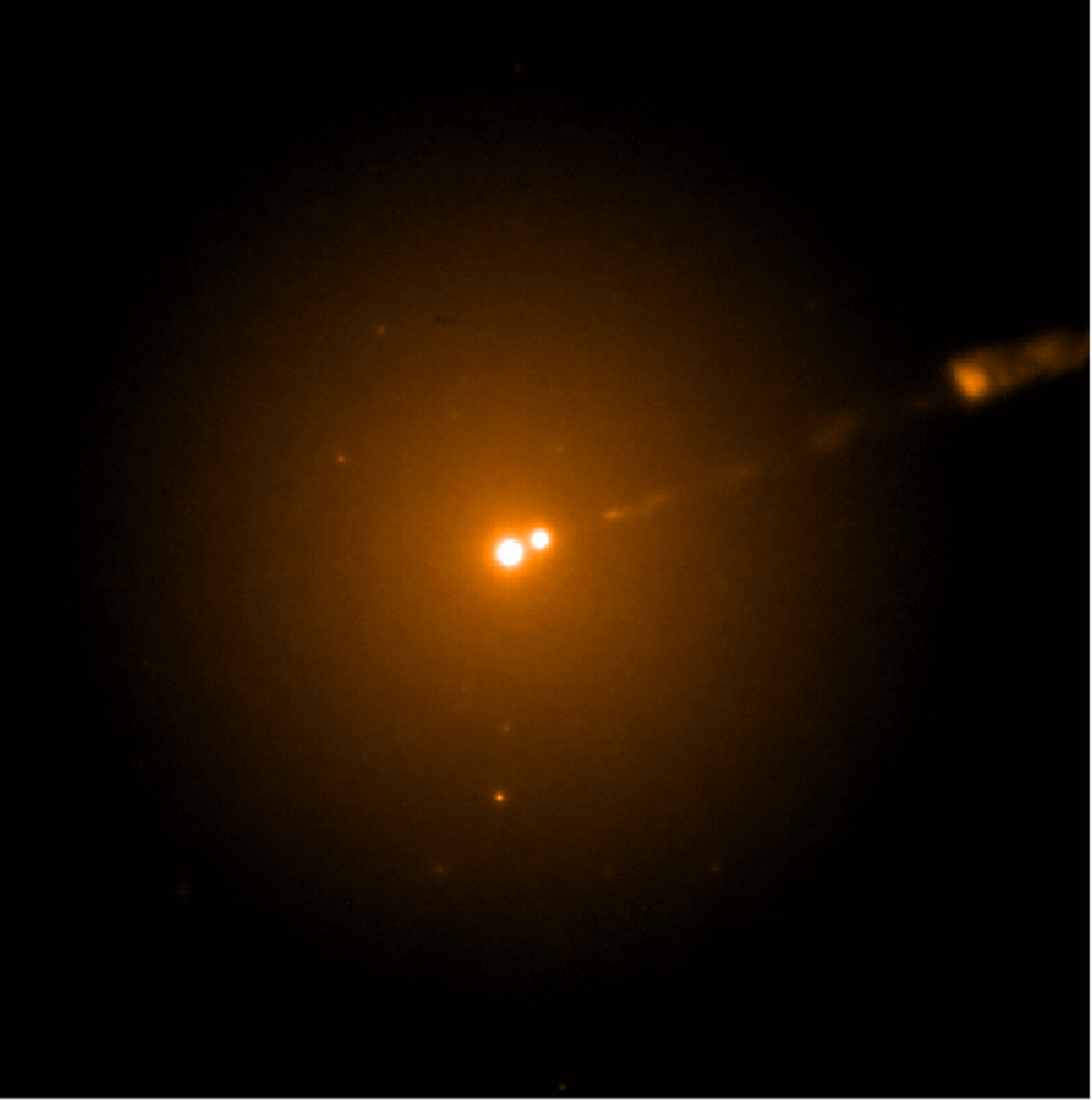}
  \caption{VLT Adaptive-Optics assisted NACO image at $2\, \rm{\micron}$ of the central kpc of M87 taken in January 2005, at the peak of M87's outburst. The brightest source at the centre is the M87 core; HST-1 is the second brightest source at 0.85 arcsec projected distance. At the time of collecting this image, HST-1 reached a maximum flux of about half that of the core.}\label{fig6}
 \end{figure*}

\subsection{Model implications and discussion}

{\bf \subsubsection{The 0.4 arcsec radius SEDs}}
The luminosity of the core as integrated from the SED model is found to be in fair agreement with that derived from the data. The result is expected given the relatively good fit of the observed SEDs. Specifically, the bolometric luminosity in the quiescent SED (integrated from the radio to the UV) is $L_{\rm obs} = 3.6 \times 10^{-6}\, L_{\rm edd} = 2.7 \times 10^{42}\, \rm{erg\,s^{-1}}$, and that from the model is $L_{\rm model} \sim 2. \times 10^{-6}\, L_{\rm edd}$. The difference is formally a factor $1.7$, which is within the range of the flux scale factor between the quiescent and the active SED (Section 3). We note that the X-ray luminosity, as compared with the total, is one order of magnitude lower, $L_{2-10\, \rm{keV}} \sim 3 \pm 1 \times 10^{41}\, \rm{erg\,s^{-1}}$.

The total jet power, $N_{\rm j}$, is found to be of the same order of magnitude in both the quiescent and the active state (Table 5). This convergence is also to be expected given the similar shape of both SEDs. The determination of $N_{\rm j}$, depends mostly --\,although not entirely\,-- on the radio data, and here the difference between both SEDs is within the errors.

Focusing on the quiescent SED, the derived jet power, which accounts for the total radiative and kinetic power, is $N_{\rm j} = 2.3 \times 10^{-6}\, L_{edd}$ (Table 5). This value is close to the total observed radiative luminosity, but this luminosity is presumably boosted by at least an order of magnitude --\,we estimate the boosting factor on the basis of a jet angle of 15 degrees and jet speed at the base of 0.5$c$--0.6$c$ \citep[][see also below]{wan09,had14}. Thus, the derived $N_{\rm j}$ comfortably accounts for the radiative power in the jet frame. The dimension at the base of the jet (radius $r_0$) is found in the model in the range of $5$ to $10\, R_{\rm g}$ for the quiescent and active SED respectively (Table 5). It is interesting to note that these values are in agreement with first estimates of the M87 event horizon from VLBI at 1.3 millimetres \citep{doe12}. The authors reported values in the $\sim 5$--$8\, R_{\rm g}$ range depending on the model fit to the VLBI data, the major uncertainty being the still limited VLBI base line coverage.

Independent estimates of the jet kinetic power in M87 have been derived from X-ray and radio observations under different model assumptions. Using arguments based on the internal pressure exerted by different knots in the jet, the inferred kinetic power is found to be $\sim 10^{44}\, \rm{erg\, s^{-1}}$ \citep[$\sim 1.4 \times 10^{-4}\, L_{\rm edd}$; e.g.][]{bic96,owe00,sta06}. The latter authors used HST-1 as a reference; hence, their estimate refers to the nearest to the core. From arguments based on the PdV work done by the jet to inflate the cavities produced in M87's X-ray halo and the regions of enhanced X-ray emission associated with shocks, estimates of the kinetic power range from $Q_{\rm edd} \sim 3.2 \times 10^{-5}$ \citep{for05} to $2.6 \times 10^{-5}$ \citep{all06} and $1.2 \times 10^{-5}$ \citep{rus13}. Similar values are inferred to balance the cooling losses from the observed X-ray luminosity, which, in the absence of a radiatively powerful core as is the case of M87, sets a lower limit to the jet mechanical power of $\gtrsim 3 \times 10^{43}\, \rm{erg s^{-1}} = 4 \times 10^{-5}\, L_{\rm edd}$ \citep{chu02}. These estimates are all subject to uncertainties, chiefly the somewhat arbitrary definition of the cavity volume and the assumption of jet power constancy, which is not the case, as illustrated by the high amplitude outbursts experienced by, for example, HST-1. They all, however, agree to within the range of $10^{-5}$--$10^{-4}\, L_{\rm edd}$. The total power $N_{\rm j}$ derived from our modelling is one to two orders of magnitude lower.

In the present modelling, the energy density in protons and electrons is assumed to be the same --\,this is not to be confused with the energy equipartition factor $k$ which is a free parameter in the model. Assuming instead the number density in protons and electrons to be equal will yield a $N_{\rm j}$ an order of magnitude higher and reach agreement with above estimates within the lower limit range. Alternatively, the higher power derived from the X-ray cavities and cooling balance may be representative of previous periods of much higher activity in M87, different from the states sampled by the present SEDs. A drawback with this argument is that by parallelism it would have to be applied to most of the nearby LLAGN for which these jet power estimates have been produced \citep[e.g.][]{all06}. Given their characteristic low luminosity and featureless spectrum \citep{fer12}, they all have to be explained by a previous period of major activity.

In spite of the uncertainties inherent to all models and power estimates, the present $N_{\rm j}$ estimate derives from the closest ever region to the core, effectively the inner $8\, \rm{pc}$, with one of the SEDs moreover tracing one of most active periods recorded for M87, the 2005 event, which all together should provide a best genuine representation of the intrinsic M87 jet power at its current state. The next more contemporaneous estimate of the jet power is that derived from the HST-1 2005 event interpreted as a shock at $68\, \rm{pc}$ distance from the core. \citet{sta06} infer a jet power of $\sim 10^{44}\, \rm{erg\,s^{-1}}$, two orders of magnitude above our $N_{\rm j}$. Assuming a jet speed at the base of $0.6\, \rm{c}$ (see references above), the 2005 outburst might represent the echo of an activity hit in M87 core more than 300 years ago, the core emission measured in the current SEDs had to be associated with quiescent epochs, the factor two stronger luminosity in the active 2005 SED being then unrelated with HST-1 outburst.

There is a robust result with regard to the derived upper limit contribution of the standard thin accretion disc to the overall energetics. This upper limit, and in turn the estimated accretion rate, is not model-dependent but imposed by the data. As illustrated in Fig.\,3, the disc contribution is somewhat constrained in the quiescent SED but unconstrained in the active one, with the jet component dominating the emission. Focusing on the quiescent SED, the inferred disc power is at most $L_{\rm disc} = 3.4 \times 10^{41}\, \rm{erg\,s^{-1}} \sim 4.6 \times 10^{-7}\, L_{\rm edd}$.

On the assumption of an efficiency of 10\%, the inferred disc power implies a strict upper limit to the accretion rate of $< 6\times 10^{-5}\, \rm{M_\odot \, yr^{-1}}$. On this basis, the radiative luminosity in the jet frame had to be boosted by a factor of 8 to 16 just to account for the observed bolometric luminosity alone, $ 2.7 \times 10^{42}\, \rm{erg\, s^{-1}}$ for the quiescent state, a factor two more for the active mode. These boosting factors are in line with our above estimate assuming a jet angle of 15 degrees. We estimate the maximum possible radiative power at the jet frame by setting our model with a jet inclination angle of 90 degrees, which yields the minimum boost possible. Integrating the SED produced by that model gives $L_{90^\circ} = 4 \times 10^{41}\, \rm{erg\,s^{-1}}$, the luminosity in the jet frame would be slightly lower and can thus be taken as a strict upper limit to the radiative power at the jet frame. This value is in line with the inferred accretion rate assuming a 10\% efficiency; however, it is an order of magnitude lower than that required to account for our inferred total jet power $N_{\rm j}$. We estimated an order of magnitude uncertainty in the determination of $N_{\rm j}$ (Section 4). Allowing for $N_{\rm j}$ to be up to a factor 10 lower to meet the accretion power will increase the difference with the kinetic power estimates quoted above to three order of magnitude.

The derived accretion rate is two to four orders of magnitude lower, respectively, than estimates based on Bondi premises, $0.01$--$0.2\, \rm{M_\odot \, yr^{-1}}$ \citep{mat03,chu02,rus15} or the jet mechanical power, $\sim 10^{-3}\, \rm{M_\odot \, yr^{-1}}$ \citep{bro15}. However, it is in line with the upper limits from Faraday rotation measurements at sub-millimetre wavelengths at the centre of M87, $\dot{M} < 9 \times 10^{-4}\, \rm{M_\odot \, yr^{-1}}$ \citep{kuo14}, and with the gas estimates from the H$\alpha$ Keplerian disc \citep[Section 1;][]{for94}, $\dot{M} \sim 10^{-8}\, \rm{M_\odot \, yr^{-1}}$ (this value is derived from the reported H$\alpha$ disc flux and 10\% efficiency). On theoretical grounds, extremely low accretion rates are inferred from hydrodynamical simulations of rotating, axisymmetric accretion flows with Bremsstrahlung cooling \citep{li13}, from magnetohydrodynamic simulations of accreting BH in which magnetically arrested accretion discs lead to efficient outflows --\,jets\,-- for high BH spins \citep{tch11}. The latter predicts BH efficiencies up to 200 per cent at peak states; the only way to get that net energy is by extracting it from the BH spin \citep{pen69,bla77}. Yet, given the low accretion rate inferred for M87, much higher efficiency, almost factor 10 higher, would be required to account for the kinetic power estimates from the X-ray cavities and cooling balance.

We finally explore whether the UV photons produced by the jet component are sufficient to account for M87's H$\alpha$ emission from its nuclear Keplerian disc: $L_{\rm H\alpha} = 1.3 \times 10^{39}\, \rm{erg\,s^{-1}}$ \citep{for94}. The ionizing Lyman continuum was integrated from the model, converted to H$\alpha$ luminosity and corrected for Galactic extinction (Section 2). Assuming a boosting factor of 10 in the observer frame, the luminosity in the jet frame is found to be $ L_{\rm H\alpha, model} = 1 \times 10^{38}\, \rm{erg\,s^{-1}}$, an order of magnitude lower than required. Accordingly, photo-ionization by the jet-core alone is insufficient to account for the observed H$\alpha$. The imbalance is not surprising; shock excitation induced by the jet propagating outward should further contribute to the gas ionization. Sources dominated by shock excitation have a characteristic LINER spectrum \citep{con97}, this being the case for M87 core and extended ionized gas spectrum.

\subsubsection{The highest angular resolution SED}
The inferred total jet power from the modelling of this SED is $N_{\rm j} = 5.2 \times 10^{-7}\, L_{\rm edd} \sim 3.8 \times 10^{41}\, \rm{erg\, s^{-1}}$, i.e. an order of magnitude lower than that derived from the 0.4 arcsec SEDs, two to three orders of magnitude lower than the jet kinetic power derived from the X-ray both cavities and cooling balance. The decrease in $N_{\rm j}$ is driven by the much lower VLBI fluxes --\,$N_{\rm j}$ is largely constrained by the cm data. However, if the VLBI cm data are optically thick (Section 3.1), $N_{\rm j}$ may in this case not be representative of the total jet power. Conversely, the 0.4 arcsec SEDs which sample the same physical region across the entire spectrum should provide a more genuine account of the total power.

We further found that by fixing the disc contribution to that derived from the 0.4 arcsec SED, the resulting model provides a qualitative good fit of the SED (Fig.\,5), in particular, the transition region from the optically thick to thin frequencies is correctly reproduced. To illustrate this effect, we added in Fig.\,5 the ALMA data to the high angular resolution SED. The ALMA data are not used in the fit but it can be seen that despite their lower resolution, they closely follow the leading trend of the model, from the cm to the highest frequencies ALMA data points. The spectral turnover occurs in the mm region, and thus we identify this spectral region as the transition point where the jet becomes optically thick. A robust result arising from this model is that the disc contribution estimated from the 0.4 arcsec SED is fully compatible with the flux limits imposed by the 0.15 arcsec aperture radius, which in turn means that our estimated disc luminosity and accretion rate are effectively constrained to the central $8\, \rm{pc}$ of M87.

 \subsubsection{Modelling conclusion: the 0.4 arcsec vs the 0.15 arcsec angular resolution SEDs}
Although it is tempting to associate this high resolution SED with the core of M87, we believe that caution should be exercised owing to the different angular resolution across the spectrum and the optically thick nature of the VLBI emission.We found that the 0.4 arcsec SED provides the most rigorous representation of the inner $8\, \rm{pc}$ jet-core of M87. The power-law like form of the compiled SEDs strictly limit the contribution of a standard thin accretion disc. This result, imposed by the data, has to be put in the context of other compelling evidence of a disc in M87: a nuclear, ionized-gas Keplerian disc perpendicular to the jet axis \citep{for94}, the thick accretion disc suggested from VLBI observations of the jet collimation at 30--100 Schwarzschild radii from the BH \citep{jun99}. On the evidence shown by the present SEDs, the signature of a standard thin accretion disc or a radiatively inefficient flow remains elusive down to $8\, \rm{pc}$ from the centre, whereas the jet is the dominant component at all scales. The jet boosting factor, estimated to be about 10, may hamper the detection of a still weaker thin disc or RIAF emitting underneath. As noted in \citet{mar05}, the base of the jets can be seen as an interface region incorporating a hot corona where the jets are launched. However, any canonical RIAF component seems to be sub-luminous in comparison. In this context, we note the results by \citet{chi02} on the nature of the UV emission of a sample of FRI radio galaxies with jets in different orientations. The UV emission is found to be of non-thermal nature in all cases regardless of jet orientation, implying that the elusive disc signature is genuine and not an effect of dimming by the jet.

\subsection{Comparison with jet+disc modelling of BHB and other low-luminosity AGN}
The jet--disc model has also been applied to other low Eddington sources including the low-hard state of BHB (\citealt{mar01,mar05,mai09}; and see others in \citealt{mar10}), to LLAGN such as M81, NGC\,4051 and Sgr~A$^*$ \citep[e.g.][]{mar01,mar08,mai11}. This section examines the parameter range inferred for M87 in the framework of those found in the former objects. For this, the compilation of model parameters in \citet{mai11} is used. The range of fractional $L_{\rm edd}$ sampled for BHB goes from $10^{-7}$--$10^{-1}$, whereas for the LLAGN the range explored so far is restricted to the lower end: $\sim 10^{-5}$ in M81, $10^{-7}$--$10^{-6}$ in M87 (see former subsection), $10^{-8}$ in Sgr~A$^*$. The disc contribution is negligible in M87 whereas in M81 it is relatively important, mostly as an additional source of Comptonized photons for the X-ray spectrum. In M87, the X-rays can, however, be fully accounted for by the Comptonized component of the jet. It should be noted that the UV to IR region, where any disc component should peak, is very well-sampled in M87 in terms of angular resolution and frequency coverage, as compared with, for example, M81, where some contribution from the galaxy stellar population could pollute the UV emission, or with Sgr~A$^*$, where the frequency coverage is limited by the enormous extinction towards the centre of our Galaxy. None of these limitations apply to M87 SEDs. The robustness of our results might prompt to a re-evaluation of the disc contribution in those other sources.

The spectral index of the particles at the post-shock region, $p$, is in the range $2.4$--$2.9$ in BHB, as in M81. In M87 it is constrained to $p \sim 3$, which is at the high end of what is expected from an initial distribution of particles accelerated by a diffusive shock process \citep[$p \sim 2.5 $; e.g.][]{hea88}. The determination of the spectral index is extremely dependent on how well the SED is sampled but also on the relative contribution of the disc and jet. The frequency sampling of M87 SED is excellent; hence, the derived spectral index is quite robust. The case of M87 confirms a prevalence of steep jet spectral index in LLAGN, equivalent relatively steep spectra are found in other LLAGN in our sample \citep{fer12}. This steep spectrum may also result from fast cooling for an originally hard accelerated spectrum ($p \sim 1.5$) as predicted in some models \citep[e.g.][]{sir14}. The turnover frequency of the spectrum is found in the sub-millimetre in M81, Sgr~A$^*$ \citep{fal00,mel01,mai09,mar08}, as well as in M87. Because of the good frequency sampling, the determination in M87 and in Sgr~A$^*$ is robust, the coincidence may hint at a similar jet structure in these sources. This possibility will be fully explored with the upcoming Event Horizon Telescope project \citep{doe12}, in which these are the two primary targets.

Finally, it is worth noting that the dominance of a jet component in M87 is in line with the known trend in BHB of becoming increasingly jet dominated at low luminosity. It is actually implied by the fundamental plane relation between radio and X-ray luminosity \citep{fen03}.

\section{Conclusions}

This paper presents quasi simultaneous high angular resolution spectral energy distributions of the LLAGN M87. Two representative SEDs of M87's core state are produced, one representing the most common, relatively quiescent mode of M87 core, and a second one representing an active phase compiled for one of the most spectacular outbursts recorded for M87, the 2005 event. Both sample the same physical scale, a radius of $32\, \rm{pc}$ from the centre, across the entire electromagnetic spectrum. In contrast with previous studies of LLAGN and of M87 in particular, the consistent spatial resolution and time simultaneity across the electromagnetic spectrum allows us to provide a firm estimate of its intrinsic core emission and accretion power.

Both SEDs have very similar shapes, the major difference being a constant flux increase in the active state by an average factor of $1.5$--$2$ from the X-rays to the IR bands. This scaling factor appears to be a characteristic parameter regulating the energy output of M87: almost the entire spectral energy distribution is modulated by this factor in a quasi-simultaneous manner whereas its shape remains unchanged. Such a variability pattern contrasts with the usual more complex ones shown by powerful AGN, and it may be a characteristic of LLAGN.

Comparing M87 core SEDs with those of powerful AGN, namely Seyfert types compiled on equivalent physical scales, and radio-loud quasars, M87 stands out by lacking the three major ingredients characterizing an AGN: \textit{i)} the blue bump signature of a thin accretion disc, \textit{ii)} the IR bump signature of dust re-emission by UV photons from the accretion disc, and \textit{iii)} the $1\, \rm{\micron}$ inflection point, a signature of the inner dust sublimation temperature. The spectrum of M87 core follows a power-law form reminiscent of that seen in BL Lac, the difference being that in the latter the high energy spectrum is usually very strong, opposite to the case of M87 \citep[e.g.][]{ghi98}.

A comparison of M87 SEDs with that of HST-1 jet knot at the time of the 2005 outburst illustrates the dramatic difference in the activity state of both sources: HST-1 with a steep flux spectrum with increasing frequency should be experiencing a particle acceleration event at current time; that M87's core shows a decreasing flux spectrum with increasing frequency is, if anything, in a post particle-acceleration cooling process, for both the sampled active and quiescent states.

The analysis of VLBI images at $5$ and $1.6\, \rm{GHz}$ shows that the effectively sampled region by these SEDs is indeed the central $8\, \rm{pc}$ radius from the BH. We thus conclude that down to the scales of a few parsecs worked by these SEDs, the characteristic spectrum of a RIAF and/or a receding standard thin disc, are not seen. Any canonical RIAF component if present is sub-luminous in comparison.

Limits on the contribution of a truncated standard accretion disc in the SEDs sets a strict upper limit of the mass accretion rate of $6\times 10^{-5}\, \rm{M_\odot \, yr^{-1}}$. This value is two to three orders of magnitude lower than estimates based on either Bondi premises or the jet mechanical power. M87 is a very active source, as demonstrated by its continuous and fast variability in X-rays, and regular powerful flares at high energies and radio monitored for the last 10--15 years. Based on the strict upper limit estimate of the accretion rate, which is entirely set by the compiled SEDs, also consistent with upper limits from the Faraday rotation, the inferred power budget just accounts for the jet power implied by our jet+disc model within the model uncertainty. The largely superior jet mechanical power inferred from the X-ray cavities and cooling balance, and HST-1 outbursts had then to be due to a previous very high activity period occurring at about 200 yr ago. A difficulty with with this explanation is that many other low luminosity AGN with extreme powerful jets would require by parallelism to have been much more active in the past to explain their underluminous spectrum. Alternatively, M87 could be extracting extra power from a spinning BH. We find, however, that the efficiency required to power M87 jets had to be at least an order of magnitude higher than those currently inferred from MHD simulations of highly spinning BHs, currently estimated at a few hundred per cent.

\section{Acknowledgements}
We are thankful to Dan Harris for providing us with the M87 NRAO radio data, flux measurements and many insightful comments on the manuscript. AP acknowledges M. Krause, E. Churazov and R. Sunyaev for discussions. SM is grateful to the University of Texas in Austin for its support, through a Tinsley Centennial Visiting Professorship. AP acknowledges financial support from the Spanish 2011 Severo Ochoa Program SEV-2011-0187, and the hospitality of the CAST group of the Ludwig Maximilians University of M\"unchen (LMU) and the Max-Planck Institute f\"ur extraterrestrische Physik where most of this work was done.\\

We dedicate this work to Dan Harris.

\bibliographystyle{mn2e}


\clearpage

\onecolumn

\begin{deluxetable}{lccc}
\tablecolumns{4}
\tablewidth{0pt}
\tablecaption{M87 core SED in quiescent phase from aperture radius $\sim 0\farcs4$. Preference is given to data collected in 2003 due to the wide frequency range covered in that year. When several observations in 2003 are available, their average is used for the modelling and shown in the SED (Fig.\,1). In the X-ray, we give the average of two \textit{Chandra} (2000 and 2002) observations with an adopted error of 50\% to account for variability. If not reference is provided, values are from this work.}\label{sed-quiescent-0.4}
\tablehead{
\colhead{Frequency (Hz)} &
\colhead{Flux (Jy)} &
\colhead{Error (Jy)} &
\colhead{Source / Date / Reference}}

\startdata
$2.42 \times 10^{26}$ & $1.6 \times 10^{-7}$  &    $-$     & $1\, \rm{TeV}$ HESS 04 \citealt{aha06} \\
$2.42 \times 10^{18}$ & $1.80 \times 10^{-8}$  & $2.3 \times 10^{-9}$   & $10\, \rm{keV}$ \textit{Chandra} 00-7 \citealt{mat03} \\
$2.42 \times 10^{18}$ & $2.80 \times 10^{-8}$  & $3.6 \times 10^{-9}$   & $10\, \rm{keV}$ \textit{Chandra} 02-7-24 \\
$2.42 \times 10^{18}$ & $2.3 \times 10^{-8}$  & $1.2 \times 10^{-8}$  & $10\, \rm{keV}$ aver. 00 \& 02, 50\% error (1)  \\
$2.42 \times 10^{17}$ & $3.10 \times 10^{-7}$  & $0.09 \times 10^{-7}$  & $1\, \rm{keV}$ \textit{Chandra} 00-7 \citealt{mat03} \\
$2.42 \times 10^{17}$ & $2.0 \times 10^{-7}$  & $0.1 \times 10^{-7}$  & $1\, \rm{keV}$ \textit{Chandra} 02-7-24 \\
$2.42 \times 10^{17}$ & $2.50 \times 10^{-7}$  & $1.25 \times 10^{-7}$  & $1\, \rm{keV}$ aver. 00 \& 02, 50\% error (1) \\
$2.06 \times 10^{15}$ & $4.73 \times 10^{-5}$ & $0.47 \times 10^{-5}$  & $1465\, \rm{\AA}$ STIS-F25SRF2 99-5-17 \\
$1.36 \times 10^{15}$ & $1.33 \times 10^{-4}$ & $0.04 \times 10^{-4}$  & F220W ACS-HRC  03-11-29 \\
$1.27 \times 10^{15}$ & $1.05 \times 10^{-4}$ & $0.03 \times 10^{-4}$  & 2360A STIS-F25QTZ 03-7-27 \\ 
$1.11 \times 10^{15}$ & $1.55 \times 10^{-4}$ & $0.03 \times 10^{-4}$  & F250W ACS-HRC 03-05-10 \\
$8.93 \times 10^{14}$ & $2.16 \times 10^{-4}$ & $0.04 \times 10^{-4}$  & F330W ACS-HRC 03-3-31 \\
$8.93 \times 10^{14}$ & $2.10 \times 10^{-4}$  & $0.04 \times 10^{-4}$  & F330W ACS-HRC 03-5-10 \\
$6.32 \times 10^{14}$ & $4.13 \times 10^{-4}$ & $0.12 \times 10^{-4}$  & F475W ACS-HRC 03-11-29 \\
$4.99 \times 10^{14}$ & $6.33 \times 10^{-4}$ & $0.63 \times 10^{-4}$  & F606W ACS-HRC 03-11-29 \\
$3.70 \times 10^{14}$ & $9.5 \times 10^{-4}$ & $1.9 \times 10^{-4}$   & F814W ACS-HRC 03-11-29 \\
$3.32 \times 10^{14}$ & $1.38 \times 10^{-3}$ & $0.1 \times 10^{-4}$   & F850LP ACS-WFC 03-1-19 \\
$2.47 \times 10^{14}$ & $2.06 \times 10^{-3}$ & $0.18 \times 10^{-3}$  & F110W NIC2 97-11-10 \\
$1.81 \times 10^{14}$ & $3.1 \times 10^{-3}$  & $0.8 \times 10^{-3}$   & F166N NIC3 99-1-16 \\ 
$1.37 \times 10^{14}$ & $3.3 \times 10^{-3}$ & $0.6 \times 10^{-3}$   & F222M NIC3 98-1-16 \\
$2.8 \times 10^{13}$  & $1.67 \times 10^{-2}$ & $9. \times 10^{-4}$  & Gemini $10.8\, \rm{\micron}$ 01-05 \citealt{per01} \\
$2.6 \times 10^{13}$  & $1.3 \times 10^{-2}$  & $2. \times 10^{-3}$  & Keck $11.7\, \rm{\micron}$ 00-1 \citealt{why04} \\
$635.0 \times 10^{9}$ & $0.43$  &	$0.09$ & ALMA 2012-6-3 \\
$350.0 \times 10^{9}$ & $0.96$  &	$0.02$ & ALMA 2012-6-3 \\
$286.0 \times 10^{9}$ & $1.28$  &	$0.02$ & ALMA 2012-6-3 \\
$252.0 \times 10^{9}$ & $1.42$  &	$0.02$ & ALMA 2012-6-3 \\
$221.0 \times 10^{9}$ & $1.63$  &	$0.03$ & ALMA 2012-6-3 \\
$108.0 \times 10^{9}$ & $1.91$  &	$0.05$ & ALMA 2012-6-3 \\
$93.7 \times 10^{9}$ & $1.82$  &	$0.06$ & ALMA 2012-6-3 \\
$22.0 \times 10^{9}$ & $2.0$  &	$0.1$  & VLA-A 03-06 \\
$15.0 \times 10^{9}$ & $2.7$  &	$0.1$  & VLA-A aver. 03-06 \& 03-08 \\
$8.4 \times 10^{9}$  & $3.15$  &	$0.16$ & VLA-A 04-12-31 \\
$8.4 \times 10^{9}$  & $3.02$  &	$0.02$ & VLA-A aver. 03-06 \& 03-08 \\
$5. \times 10^{9}$  & $3.10$  &	$0.06$ & VLA-A 99-09 \citealt{nag01} \\
\enddata

\end{deluxetable}

\begin{deluxetable}{lccc}
\tablecolumns{4}
\tablewidth{0pt}
\tablecaption{M87 core SED from 2005 --\,active phase\,-- extracted from a $0\farcs4$ aperture radius. ALL values are from this work.}\label{sed-active-0.4}
\tablehead{
\colhead{Frequency (Hz)} &
\colhead{Flux (Jy)} &
\colhead{Error (Jy)} &
\colhead{Source / Date / Reference}}

\startdata
$1.36 \times 10^{15}$  & $2.08 \times 10^{-4}$  & $0.06 \times 10^{-4}$   &   F220W ACS-HRC 05-5-9 \\ 
$1.10 \times 10^{15}$  & $3.0 \times 10^{-4}$   & $0.2 \times 10^{-4}$   &   F250W ACS-HRC 05-5-9 \\
$8.93 \times 10^{14}$  & $4.2 \times 10^{-4}$  & $0.2 \times 10^{-4}$    &   F330W ACS-HRC 05-5-9 \\
$6.32 \times 10^{14}$  & $7.61 \times 10^{-4}$  & $0.15 \times 10^{-4}$   &   F475W ACS-HRC 05-5-9 \\
$5.0 \times 10^{14}$   & $1.02 \times 10^{-3}$  & $0.13 \times 10^{-3}$   &   F606W ACS-HRC 05-5-9 	 \\
$5.0 \times 10^{14}$   & $9.52 \times 10^{-4}$  & $0.95 \times 10^{-4}$   &   F606W ACS-HRC 05-6-22 \\ 
$3.70 \times 10^{14}$  & $1.61 \times 10^{-3}$  & $0.24 \times 10^{-3}$   &   F814W ACS-HRC 05-5-9 \\
$2.46 \times 10^{14}$  & $1.6 \times 10^{-3}$   & $0.7 \times 10^{-3}$    &   \textit{J}-band VLT-NACO 05-01 \\
$1.37 \times 10^{14}$  & $2.67 \times 10^{-3}$  & $0.55 \times 10^{-3}$   &   \textit{K}-band VLT-NACO 05-01 \\
$3.7 \times 10^{13}$   & $1.0 \times 10^{-2}$    & $0.5 \times 10^{-2}$    &   $8\, \rm{\micron}$ Subaru-spec. 05-04 \\
$2.7 \times 10^{13}$   & $1.4 \times 10^{-2}$   & $1.4 \times 10^{-2}$     &   $11\, \rm{\micron}$ Subaru-spec. 05-04 \\
$22.0 \times 10^{9}$   & $2.75$               & $0.14$	                &   VLA-A 05-5-3 \\
$15.0 \times 10^{9}$   & $3.08$               & $0.15$                  &   VLA-B 05-5-3 \\
\enddata

\end{deluxetable}

\begin{deluxetable}{lccc}
\tablecolumns{4}
\tablewidth{0pt}
\tablecaption{SED from HST-1 in 2005 in an from an aperture radius of $0\farcs4$. If not reference is provided, values are from this work.}\label{sed-active-0.4}
\tablehead{
\colhead{Frequency (Hz)} &
\colhead{Flux (Jy)} &
\colhead{Error (Jy)} &
\colhead{Source / Date / Reference}}

\startdata
$4.84  \times 10^{17}$ &  $1.07 \times 10^{-6}$ &   $0.25 \times 10^{-6}$  & \textit{Chandra} $2\, \rm{keV}$,  05-2-9, \citealt{har06} \\
$1.36  \times 10^{15}$ &  $7.50 \times 10^{-4}$  &   $0.37 \times 10^{-4}$  & ACS-HRC-F250W 05-2-9 \\
$5.00  \times 10^{14}$ &  $1.01 \times 10^{-3}$  &   $0.10 \times 10^{-3}$   & ACS-HRC-F606W 05-2-9  \\
$2.46  \times 10^{14}$ &  $1.00 \times 10^{-3}$  &   $0.19 \times 10^{-3}$  & NACO-J-band, 05-1-20 \\
$1.37  \times 10^{14}$ &  $1.16 \times 10^{-3}$  &   $0.15 \times 10^{-3}$  & NACO-K-band 05-1-20 \\
$22.48 \times 10^{9}$  &  $0.067$               &   $0.010$                 & VLA-A 05-01 \\
$14.96 \times 10^{9}$  &  $0.082$               &   $0.010$	           & VLA-A 05-01 \\
$8.43  \times 10^{9}$  &  $0.119$               &   $0.003$                & VLA-A 05-01 \\
\enddata

\end{deluxetable}

\begin{deluxetable}{lccc}
\tablecolumns{4}
\tablewidth{0pt}
\tablecaption{Highest spatial resolution SED of M87 core in quiescent phase, i.e.\ outside the flare periods. When multiple radio observations are available, the average is provided. If no reference is provided, values are from this work.}\label{SED-quiescent-0.2}
\tablehead{
\colhead{Frequency (Hz)} &
\colhead{Flux (Jy)} &
\colhead{Error (Jy)} &
\colhead{Source / Date / Reference / Aperture}}

\startdata
$2.06 \times 10^{15}$  &   $4.14 \times 10^{-5}$ &  $4.1\times 10^{-6}$ & $1465\, \rm{\AA}$ STIS-F25SRF2 99-5-17 $r = 0\farcs12$ \\
$1.36 \times 10^{15}$  &	$1.08 \times 10^{-4}$ &  $0.02 \times 10^{-4}$ & F220W ACS-HRC 03-11-29 $r = 0\farcs13$  \\
$1.27 \times 10^{15}$  &   $1.81 \times 10^{-4}$ &  $2. \times 10^{-6}$ & 2360A STIS-F25QTZ 01-7-30  $r = 0\farcs12$ \\
$1.10 \times 10^{15}$  &	$1.28 \times 10^{-4}$ &  $0.01 \times 10^{-4}$ & F250W ACS-HRC 03-05-10 $r = 0\farcs13$ \\
$8.93 \times 10^{14}$  &	$1.79 \times 10^{-4}$ &  $1.4 \times 10^{-5}$  & F330W ACS-HRC 03-3-31  $r = 0\farcs13$ \\
$6.32 \times 10^{14}$  &   $4.13 \times 10^{-4}$ &  $0.54 \times 10^{-4}$ & F475W ACS-HRC 03-11-29 $r = 0\farcs13$ \\
$4.99 \times 10^{14}$  &	$4.24 \times 10^{-4}$ &  $0.24 \times 10^{-4}$ & F606W ACS-HRC 03-11-29 $r = 0\farcs13$ \\  
$3.70 \times 10^{14}$  &	$6.34 \times 10^{-4}$ &  $1.26 \times 10^{-4}$ & F814W ACS-HRC 03-11-29 $r = 0\farcs15$ \\  
$3.32 \times 10^{14}$  &   $1.30 \times 10^{-3}$  &  $0.14 \times 10^{-3}$ & F850LP ACS-WF 03-1-19  $r = 0\farcs14$ \\
$2.47 \times 10^{14}$  &   $1.61 \times 10^{-3}$ &  $0.16 \times 10^{-3}$ & F110W NIC2 97-11-10 $r = 0\farcs15$   \\
$2.3 \times 10^{11}$	  &   $0.98$	         &  $0.04$	   & VLBI $1.3\, \rm{mm}$  2009, \citealt{doe12},  \textsc{fwhm} = $40 \pm 1.8\, \rm{mas}$ \\
$1.00 \times 10^{11}$  &	$5.00 \times 10^{-1}$ &  $0.1$     & VLBI $3\, \rm{mm}$ sometime in 95-12 -- 96-4, \citealt{lon98} \\
$86.0 \times 10^{9}$	  &  	$0.16$               &  $0.07$    & VLBI 2001, \citealt{lee08}, beam = $198 \times 78\, \rm{\mu as^2}$ \\
$22.0 \times 10^{9}$	  &	$0.35$               &  $0.06$    & VLBI 1992 aveg. 2 obs., \citealt{jun95}, beam = $1.15 \times 0.14\, \rm{mas^2}$ \\
$8.4 \times 10^{9}$	  &     $0.52$               &  $0.05$    & VLBI sometime in 1981-1984, \citealt{mor86} beam $<1\, \rm{mas}$ \\
$2.3 \times 10^{9}$    &	$0.63$               &  $0.08$    & VLBI aveg. obsv. 1981-1984, \citealt{mor88},  beam $<3\, \rm{mas}$ \\
$1.6 \times 10^{9}$    &	$0.364$	             &   $-$      & VLBI 84-4, \citealt{gio90},  beam = $4 \times 4\, \rm{mas^2}$ \\
\enddata

\end{deluxetable}

\begin{deluxetable}{lccc}
\tablecolumns{4}
\tablewidth{0pt}
\tablecaption{Model results: quiescent and active M87 SEDs with aperture radius $\sim 0\farcs4$. Values with $^*$ are fixed in the fit (see text).}\label{sed-model}
\tablehead{
\colhead{Parameter} &
\colhead{Quiescent} &
\colhead{Active}}

\startdata
{Pre-shock Maxwell electron distribution model} \\
\\
\hline
\hline\\[-0.3cm]
$N_{\rm j}$ jet normalization & $2.3 \times 10^{-6}\, L_{\rm edd}$ & $2.5 \times 10^{-6}\, L_{\rm edd}$ \\ 
$T_{\rm e}$ pre-shock e$^-$ temp.  & $2.9 \times 10^{11}\, \rm{K}$ & $3.0 \times 10^{11}\, \rm{K}$ \\  
$k$ equipartition & 0.3 & 0.3	 \\
$r_0$ jet base radius & $5.3\, R_{\rm g}$ & $10.0\, R_{\rm g}$ \\ 
$h_0/r_0$ jet base height-to-radius & 2.0 & 2.3	\\
$z_{\rm acc}$ starting point for acceleration in jet & $10\, R_{\rm g}$ & $8\, R_{\rm g}$\\
$p$ spectral index post-shock e$^-$ & 3.0 & 2.9  \\ 
plfrac fraction non-thermal e$^-$ at post-shock & 0.6$^*$ & 0.6$^*$ \\ 
$\log(z_{\rm max})$ & $10^{19}\, \rm{cm}^*$ & $10^{19}\, \rm{cm}^*$ & \\
$r_{\rm in}$ inner radius AD & $5\, R_{\rm g}$ & $5\, R_{\rm g}$ & \\
$r_{\rm out}$ outer radius AD & $200\, R_{\rm g}^*$ & $200\, R_{\rm g}^*$ & \\ 
$T_{\rm in}$ innermost AD Temp. & $2000\, \rm{K}$ & $1500\, \rm{K}$ \\ 
\enddata

\end{deluxetable}

\begin{deluxetable}{lccc}
\tablecolumns{2}
\tablewidth{0pt}
\tablecaption{Model result: M87 SED with the highest angular resolution, and quiescent mode. Aperture radius $\sim 0\farcs15$ in the optical--UV, in the model taken as upper limits, VLBI mas in the cm -- mm range. Values with $^*$ are fixed in the fit.}\label{sed-model-VLBI}
\tablehead{
\colhead{Parameter} &
\colhead{Quiescent}}

\startdata
{Pre-shock Maxwell electron distribution model} \\
\\
\hline
\hline\\[-0.3cm]
$N_{\rm j}$ jet normalization & $5.2 \times 10^{-7}\, L_{\rm edd}$ \\  
$T_{\rm e}$ pre-shock e$^-$ temp.  & $4.5 \times 10^{11}\, \rm{K}$ \\
$k$ equipartition & 1.1	 \\
$r_0$ jet base radius & $3\, R_{\rm g}$ \\ 
$h_0/r_0$ jet base height-to-radius & 10 \\
$z_{\rm acc}$ starting point for acceleration in jet & $50\, R_{\rm g}$  \\
$p$ spectral index post-shock e$^-$ & 2.5$^*$  \\
plfrac fraction non-thermal e$^-$ at post-shock & 0.6$^*$  \\
$\log(z_{\rm max})$ & $10^{19}\, \rm{cm}^*$\\
$r_{\rm in}$ inner radius AD & $5\, R_{\rm g}$ \\
$r_{\rm out}$ outer radius AD & $200\, R_{\rm g}^*$  \\
$T_{\rm in}$ innermost AD Temp. & $1000\, \rm{K}$  \\
\enddata

\end{deluxetable}

\label{lastpage}
\end{document}